\documentstyle[preprint,aps]{revtex}
\begin{document}
\draft
\preprint{DESY 94-199}
\title{
The Geroch group in the Ashtekar formulation}
\author{Shun'ya MIZOGUCHI
\footnote{Fellow of the Alexander von Humboldt Foundation}
\footnote{Electronic address: mizoguch@x4u2.desy.de
}}
\address
{II. Institut f\"{u}r Theoretische Physik,
            Universit\"{a}t Hamburg \\
            Luruper Chaussee 149, 22761 Hamburg, B.R.D.
}
\date{November 4, 1994}
\maketitle
\begin{abstract}
We study the Geroch group in the framework of the Ashtekar
formulation. In the case of the one-Killing-vector reduction,
it turns out that the third column of the Ashtekar connection
is essentially the gradient of the Ernst potential, which
implies that the both quantities are based on the ``same''
complexification. In the two-Killing-vector reduction, we
demonstrate Ehlers' and Matzner-Misner's SL(2,R) symmetries,
respectively, by constructing two sets of canonical variables
that realize either of the symmetries canonically, in terms
of the Ashtekar variables. The conserved charges associated
with these symmetries are explicitly obtained. We show that
the gl(2,R) loop algebra constructed previously in the loop
representation is not the Lie algebra of the Geroch group
itself. We also point out that the recent argument on the
equivalence to a chiral model is based on a gauge-choice
which cannot be achieved generically.  \\
PACS number: 04.20.Fy
\end{abstract}

\narrowtext

\section{Introduction}
The Geroch group has been known, since his discovery in 1971,
as a symmetry group acting on the solutions of Einstein's
equation \cite{Geroch}. Developing the early day's result
on the (actually two types of) SL(2,R) moduli in the presence
of one or two Killing vector fields \cite{Ehlers,MM}, he
showed that the symmetry of the
solutions is enlarged to infinite-dimensional in consequence
of the interplay between two different SL(2,R) symmetries.
This result was further developed as solution-generating
techniques in terms of general relativity, and several
approaches to generate solutions have been subsequently
evolved \cite{Esol}. Among them was shown the integrability
of the Ernst equation \cite{Ernst}, which determines the
solutions of Einstein's equation for stationary axisymmetric
spacetime, by explicitly constructing the Lax pair for this
equation \cite{BS,Maison}.

Some years later, particle physicists also became interested
in the Geroch group, based on the recognition that the
emergence of the extra symmetry can be realized in the same
context as that of hidden symmetries \cite{CremmerJulia} in
the Kaluza-Klein reduction of supergravity
theories. This was motivated by the work of B.Julia
\cite{Julia}, who showed in 1980 that the Lie algebra
of the Geroch group is in fact the
${\rm\widehat{sl}(2,R)}$ affine Kac-Moody algebra.
Moreover, he pointed out the existence of the non-zero
central term of this algebra even at the classical level,
which acts as a constant rescaling on the conformal factor
of the zweibein in the resulting two-dimensional field theory.
The group-theoretical structure was further elaborated \cite{BM}
in connection with the non-linear sigma model, and the
generalization to the Einstein-Maxwell theory was discussed
\cite{BMG}. The reduction of supergravity theory was also
explored as an integrable system \cite{Nicolai_integrable}.
An evidence of a symmetry of a hyperbolic algebra was revealed
in the reduction to one dimension \cite{Nicolai_hyperbolic}.

In canonical gravity, on the other hand, an important
breakthrough was brought about by A. Ashtekar in 1986
\cite{Ashtekar}. He found a new set of canonical variables,
in terms of which a drastic simplification occured in the
canonical constraints. Making use of the new variables allows
us to write them in a polynomial form, and what is more,
a large class of solutions for the quantum constraints can
be found by introducing the self-dual  and
the loop representation
\cite{Ashtekar} - \cite{solution},
although the interpretation of these solutions still remains
to be a difficult problem. This formalism has been subsequently
applied to spacetime with one or more Killing vector fields
(``mini-superspace'')
\cite{Ashtekar_spherical} - \cite{Manojlovic}
(\cite{Ashtekar_bib} is an exhaustive reference list for the
literature related to the Ashtekar variable.).

One of the notable features of the model with Killing-field
isometries is the existence of physical observables in the sense
of Dirac. Associated with the hidden symmetry arising through
the reduction to lower dimensions, one may always have the
symmetry charges, which by definition weakly commute with all
the constraints.  This would be helpful for better understanding
of both classical and quantum gravity, since, no such
functional is known in the ordinary four-dimensional gravity
theory, except for the constraints themselves.
The cosmological model with a closed space manifold which allows
two commuting Killing vector fields has been known as the Gowdy
model \cite{Gowdy}. The Ashtekar formulation was applied to the
three-torus topology model, and a set of operators which forms
a GL(2,R) loop algebra are constructed in the loop
representation \cite{HS}. It was conjectured that this loop
algebra $\widehat{\rm{gl}}(2,R)$ would be related to the Geroch
group. One of the aims of this paper is to clarify this point.

We first consider the one-Killing-vector reduction,
and show that the complex Ernst
potential is a ``natural'' variable in the Ashtekar formulation.
One of the distinguished properties of the Ashtekar connection
is its being a {\em complex} canonical variable. This requires
the reality condition in order to recover the ordinary general
relativity theory. On the other hand,
the complex nature of the Ernst potential is originated from
the complex structure of the target space of the coset
non-linear sigma model (the upper-half plane in our case), which
arises through the dimensional reduction of Einstein's action.
Rather unexpectedly, it turns out that $A_{z3}$ and $iA_{z\alpha}$
$(\alpha=1,2)$ are nothing but its gradient. Moreover,
the upper-indexed elements $A^{\mu}_{~\alpha}$ and $A^{\mu}_{~3}$
are {\em real} (or {\em pure imaginary}) function
(not a ``real-valued'' function, of course) of the
Ernst potential.  In this sense one
may say that these two kinds of complex variables are based on
the ``same'' complexification. Another reason why this relation
is non-trivial is that, on the course of reduction, the Ernst
potential is defined only after the duality relation is invoked;
the Ashtekar connections are so constructed that this step may
be already included in its definition. As a consequence,
$A_{za}$ $(a=1,2,3)$ transform in a simple manner under
Ehlers' SL(2,R) transformation, known as one of two SL(2,R)
subgroups whose Lie algebras generate the whole
$\widehat{\rm{sl}}(2,R)$ through the Serre relation.

We then go further to the case of two-Killing-vector reduction,
and examine how Matzner-Misner's SL(2,R), which is the other
SL(2,R), is seen in this scheme. We will see that the GL(2,R)
charges in ref.\cite{HS} act as a product of this SL(2,R) and
the center of the Geroch group. We will also show, however, that
the GL(2,R) loop algebra constructed in the loop representation
does not contain Ehlers' SL(2,R). Therefore it does not coincide
with the Lie algebra of the Geroch group itself, but is something
else. To realize Ehlers' symmetry canonically, we are forced to
work with canonical variables obtained through a non-local
canonical transformation from the original ones.
This is in some sense expected, since a similar
difficulty has been known for a long time in a chiral model,
which is a much simpler system than the present one, when one
realizes canonically the non-local Kac-Moody symmetry
\cite{chiral}.

The plan of this paper is as follows.
In Sec.II we review the two different SL(2,R) symmetries
of the reduced system which allows the presence of two commuting
Killing vectors in spacetime. In Sec.III we consider
the reduction from four dimensions to three. First we describe
the general settings for the U(1) symmetric spacetime in
Subsec.III.A, and comment on the integrality condition
considered in ref.\cite{Moncrief}.
The relation between the Ashtekar connection and
the Ernst potential is revealed in Subsec.III.B.
We prove Ehlers' SL(2,R) symmetry in the framework of the
Ashtekar formulation in Subsec.III.C, and derive the conserved
charges in Subsec.III.D. Sec.IV is devoted to the study of
the reduced model from four dimensions to two. Matzner-Misner's
SL(2,R) symmetry is demonstrated and the associated conserved
charges are obtained in Subsec.IV.A and IV.B, respectively.
In Subsec.IV.C we show that the GL(2,R) charges in ref.\cite{HS}
act as a product group of Matzner-Misner's SL(2,R) and the
center of the Geroch group. In Subsec.IV.D we examine whether
or not the loop algebra of ref.\cite{HS} includes Ehlers' SL(2,R),
and see that it does not. Finally in Sec.V, we conclude our
result, and comment on the recent argument on the equivalence
to a chiral model \cite{H}. We point out that the resulting
linear system for the flat-space SL(2,R) chiral model
is a consequence of a gauge-choice which can not be
achieved generically.

In this paper we have to group the spacetime and the Lorentz
indices in varieties of ways. Throughout the paper we use
the following notations. The capital $M,N,\ldots$ stand for
the four-dimensional spacetime indices $\{t,x,y,z\}$, and
$m,n,\ldots$ for space indices $\{x,y,z\}$.
In the one-Killing reduction, $m',n',\ldots$ represent the
reduced three-dimensional spacetime indices $\{t,x,y\}$
and $\mu,\nu,\ldots$ do the two-dimensional space indices
$\{x,y\}$, while $z$ is taken as the direction along
the Killing vector.
In the two-Killing reduction, $\tilde{m},\tilde{n},\ldots$
range over the reduced two-dimensional spacetime indices
$\{t,x\}$, while $\bar{m},\bar{n},\ldots$ are used for
the ``compactified'' coordinates $\{y,z\}$. Correspondingly,
the internal Lorentz indices
$A,B,\ldots~~$, $a,b,\ldots~~$,
$a',b'\ldots~~$, $\alpha,\beta,\ldots~~$,
$\tilde{a},\tilde{b},\ldots~~$,
and $\bar{a},\bar{b},\ldots~~$ run over
$\{0,1,2,3\}$, $\{1,2,3\}$, $\{0,1,2\}$, $\{1,2\}$,
$\{0,1\}$ and $\{2,3\}$, respectively.
We will sometimes repeat this definition of those indices
if needed in the subsequent sections. We take the signature
of the metric as $(-+++)$. The Levi-Civita
anti-symmetric tensors $\epsilon_{abc}$ and
$\epsilon_{a'b'c'}$ are so defined that $\epsilon_{123}=+1$
and $\epsilon_{012}=+1$, respectively. We restrict
ourselves only to the case in which all the Killing
vector fields are space-like in this paper.

\section{Ehlers' and Matzner-Misner's SL(2,R)}
In this section we review the Killing-vector reduction
in the Lagrangian formulation, and explain the basic two
SL(2,R) symmetries arising as a result of the reduction.
Let us first consider the reduction from four dimensions
to three by introducing a Killing vector field along the
$z$ axis. As usual in the Kaluza-Klein theory
\cite{KaluzaKlein}, we start from the following
four-dimensional metric
\begin{equation}
g_{MN}=\left(
\begin{array}{cc}
\Delta^{-1}g^{(3)}_{m'n'}+\Delta B_{m'}B_{n'}&\Delta B_{m'}\\
\Delta B_{n'}&\Delta
\end{array}
\right),
\end{equation}
where all the components are assumed to be independent
of the $z$-coordinate. This metric can be achieved by
taking the vierbein as
\begin{equation}
E_{M}^{~A}=\left(
\begin{array}{cc}
\Delta^{-\frac{1}{2}}f_{m'}^{~a'}&\Delta^{\frac{1}{2}} B_{m'}\\
0&\Delta^{\frac{1}{2}}
\end{array}
\right),\label{Esect2}
\end{equation}
where
$g^{(3)}_{m'n'}
=f_{m'}^{~a'}\eta_{a'b'}f^{b'}_{~~n'}$.
The Lagrangian is reduced to up to a total derivative
\begin{eqnarray}
{\cal L}&=&
\sqrt{-g}R\nonumber\\
&=&\sqrt{-g^{(3)}}\left[
R^{(3)}-\frac{1}{4}\Delta^2F_{m'n'}F^{m'n'}
-\frac{1}{2}g^{(3)m'n'}\Delta^{-2}
\partial_{m'}\Delta\partial_{n'}\Delta
\right],\label{L}
\end{eqnarray}
where $F_{m'n'}=\partial_{m'}B_{n'}-\partial_{n'}B_{m'}$
and $F^{m'n'}=g^{(3)m'k'}g^{(3)n'l'}F_{k'l'}$. We would
like to treat $F^{m'n'}$ as an independent field. To this end
we add the following term to the Lagrangian
\begin{equation}
{\cal L}'={\cal L}-\frac{1}{2}B\cdot\sqrt{-g^{(3)}}
\epsilon^{m'n'k'}\partial_{k'}F_{m'n'}. \label{L'}
\end{equation}
Here $B$ is the Lagrange multiplier, and
$\frac{1}{2}$ is inserted for convenience. This term guarantees
that $F^{m'n'}$ is locally a rotation. The equation of motion
of $F^{m'n'}$ then becomes
\newcounter{duality}
\setcounter{duality}{\theequation}
\begin{equation}
\Delta^2F^{m'n'}=\epsilon^{m'n'k'}\partial_{k'}B.\label{duality}
\end{equation}
Substituting (\ref{duality}) into (\ref{L'}), we obtain the
SL(2,R)/U(1) coset non-linear sigma model Lagrangian
\begin{equation}
{\cal L}'
=\sqrt{-g^{(3)}}\left[
R^{(3)}
-\frac{1}{2}g^{(3)m'n'}\Delta^{-2}
(\partial_{m'}B\partial_{n'}B
+\partial_{m'}\Delta\partial_{n'}\Delta)
\right].
\end{equation}
Setting $Z^{(E)}=B+i\Delta$, this action is invariant under
\begin{equation}
Z^{(E)} \rightarrow \frac{aZ^{(E)}+b}{cZ^{(E)}+d}
\end{equation}
for any real numbers $a,b,c,d$. Since the simultaneous
scaling on them obviously results in the same transformation, we
may impose $ad-bc=1$. Naming after the seminal work of J. Ehlers
\cite{Ehlers}, we will call this symmetry ``Ehlers' SL(2,R)''.
$Z^{(E)}$ is related to the so called Ernst
potential ${\cal E}$ \cite{Ernst} by
${\cal E}=i\overline{Z^{(E)}}$.

It turns out that another SL(2,R) symmetry arises if we further
reduce the spacetime dimension from three to two by introducing
an additional Killing vector field along the $y$ axis. Roughly
speaking, this is a symmetry of rotating in the $yz$-plane.
$f_{m'}^{~a'}$ is now assumed to be in the form
\begin{equation}
f_{m'}^{a'}=
\left(
\begin{array}{cc}
f_{\tilde{m}}^{~\tilde{a}}&\rho A_{m'}\\
0&\rho
\end{array}
\right)=
\left(
\begin{array}{cc}
\lambda \delta_{\tilde{m}}^{~\tilde{a}}&\rho A_{\tilde{m}}\\
0&\rho
\end{array}
\right).\label{fsect2}
\end{equation}
Here we have taken the conformal gauge for the zweibein of
the reduced two-dimensional field theory.
It was shown by R. Geroch that, if one would like to have
infinite-dimensional symmetry, one must assume some two
constants to vanish \cite{Geroch}. The easiest way to satisfy
this requirement is to take \cite{Julia}
\begin{equation}
A_{\tilde{m}}=B_{\tilde{m}}=0. \label{A=B=0}
\end{equation}
This means that the four-dimensional metric is assumed to be
in a block diagonal form consisting of $g_{\tilde{m}\tilde{n}}$
and $g_{\bar{m}\bar{n}}$. Evidently, it is essential that
the metric can be recasted in this form only by such a
diffeomorphism that keeps $\frac{\partial}{\partial y}$ and
$\frac{\partial}{\partial z}$ being commuting Killing vectors.
It is also clear that the Killing vectors are hyper-surface
orthogonal.
By this choice the three-dimensional
Lagrangian (\ref{L'}) becomes
\begin{eqnarray}
{\cal L'}&=&-\frac{1}{2}\rho\eta^{\tilde{m}\tilde{n}}
\left[
-4\partial_{\tilde{m}}\log\rho
\partial_{\tilde{n}}
\log\lambda
+\frac{1}{\Delta^2}\left\{
\partial_{\tilde{m}}B\partial_{\tilde{n}}B
+\partial_{\tilde{m}}\Delta
\partial_{\tilde{n}}\Delta
\right\}
\right]
.\label{L_E}
\end{eqnarray}

We may, on the other hand, perform the dimensional reduction
from four to two directly. Using (\ref{Esect2}), (\ref{fsect2})
and (\ref{A=B=0}), the Lagrangian (\ref{L}) is simplified to
\begin{eqnarray}
{\cal L}&=&-\frac{1}{2}\rho\eta^{\tilde{m}\tilde{n}}
\left[
-4\partial_{\tilde{m}}\log\rho
\partial_{\tilde{n}}
\log(\lambda\Delta^{-\frac{1}{2}}\rho^{\frac{1}{4}})
+\frac{\Delta^2}{\rho^2}\left\{
\partial_{\tilde{m}}B_y\partial_{\tilde{n}}B_y
+\partial_{\tilde{m}}\left(\frac{\rho}{\Delta}\right)
\partial_{\tilde{n}}\left(\frac{\rho}{\Delta}\right)
\right\}
\right].\label{L_MM}
\end{eqnarray}
In terms of the variable $Z^{(MM)}=B_y+i\frac{\rho}{\Delta}$, the
SL(2,R) transformation is expressed in this case
\begin{equation}
Z^{(MM)} \rightarrow \frac{aZ^{(MM)}+b}{cZ^{(MM)}+d},
\end{equation}
under which (\ref{L_MM}) is manifestly invariant. Following
ref.\cite{Julia}, we call this ``Matzner-Misner's SL(2,R)''.
Clearly the two Lagrangians (\ref{L_E}) and (\ref{L_MM}) are
made completely identical by the transformation
\begin{equation}
B\leftrightarrow B_y,~~\Delta\leftrightarrow\frac{\rho}{\Delta},
{}~~\lambda\leftrightarrow
\lambda\Delta^{-\frac{1}{2}}\rho^{\frac{1}{4}},~~
\rho\leftrightarrow\rho, \label{KN}
\end{equation}
which was found by D. Kramer and G. Neugebauer
\cite{KramerNeugebauer}.

R. Geroch noticed that the infinitesimal transformations
of these two SL(2,R) are not commutative on the solution of
Einstein's equation, but generate infinitely many different
solutions by their successive applications \cite{Geroch}.
In fact, this is isomorphic to the affine
$\widehat{\rm sl}(2,R)$ algebra \cite{Julia} (See also
Subsec.IV.C for further explanations.).
We will study in the subsequent sections the structure of the
realization of these groups in the Ashtekar formulation.

\section{Kaluza-Klein reduction to three dimensions}
\subsection{U(1) symmetric spacetime}

As seen in the previous section, the first symmetry, Ehlers'
SL(2,R), already shows up at the stage of the reduction
from four dimensions to three. Let us examine in this section
how this is seen in the Ashtekar formulation.
Although we would like to discuss its {\it local}
transformation property in this paper (since even this does not
seem to have been studied in detail before), we begin with
describing a slightly more general setting for the topology of
our spacetime that admits one Killing vector field.
This allows us to fix our notations and to comment on Moncrief's
integrality condition \cite{Moncrief}.

Our starting point is that we assume
the spacetime to be a direct product of a total space of a U(1)
principal bundle $\Sigma$ and time ${\rm R}$. The base
manifold $\widetilde{\Sigma}\sim\Sigma/{\rm U(1)}$ is assumed to
be a compact, connected and orientable two-dimensional manifold.
The fiber of the bundle is topologically ${\rm S}^1$, and the
Killing vector field assumed to exist is tangent to the fiber.
Geroch's fundamental requirement for the reduction to three
dimensions is thus satisfied by this U(1) gauge symmetry (of the
bundle). When we discuss the reduction to two dimensions
in Sec.\ref{4to2}, we will consider another Killing vector field
on the base manifold $\widetilde{\Sigma}$ in addition to the one
above.

We next introduce the U(1)-adapted coordinate $(t,x,y,z)$. Let
$t$ represent time and $(x,y)$ be a local coordinate system of
$\widetilde{\Sigma}$ on each local patch. $z$ is a coordinate of
the fiber so normalized that the Killing vector field is written as
$\frac{\partial}{\partial z}$, and $0\leq z\leq 2\pi$. This means
that all derivatives with respect to $z$ are zero for any field
that appears in the present model.

As usual in the ADM formalism \cite{ADM},
we take a vector normal to the Cauchy
surface $\Sigma$ at each point with respect to the given metric
$g_{MN}$, $M,N=t,x,y,z$. This induces a three-dimensional metric
$h_{mn}$, $m,n=x,y,z$ on $\Sigma$. $h_{mn}$ can be further
decomposed
orthogonally with respect to $\frac{\partial}{\partial z}$, which
induces a two-dimensional metric $h'_{\mu\nu}$, $\mu,\nu=x,y$ on
$\widetilde\Sigma$.
The spacetime metric is then written as
\begin{eqnarray}
ds^2&=&g_{MN}dx^M dx^N\nonumber\\
&=&-N^2dt^2+h_{mn}(dx^m+N_{(0)}^mdt)(dx^n+N_{(0)}^ndt)\nonumber\\
&=&-N^2dt^2
+\left(h_{\mu\nu}-\frac{h_{\mu z}h_{\nu z}}{h_{zz}}\right)
(dx^{\mu}+N_{(0)}^{\mu}dt)(dx^{\nu}+N_{(0)}^{\nu}dt)
\nonumber\\&&
+h_{zz}\left\{dz+N_{(0)}^zdt +\frac{h_{\mu z}}{h_{zz}}
(dx^{\mu}+N_{(0)}^{\mu}dt)\right\}^2.
\end{eqnarray}
We set
\begin{eqnarray}
h_{zz}&=&\Delta,\nonumber\\
\frac{h_{\mu z}}{h_{zz}}&=&B_{\mu},\nonumber\\
N_{(0)}^z+\frac{h_{\mu z}}{h_{zz}}
N_{(0)}^{\mu}&=&N'^3,\nonumber\\
(N)^2&=&\Delta^{-1}(N')^2,\nonumber\\
h_{\mu\nu}-\frac{h_{\mu z}h_{\nu z}}{h_{zz}}
&=&\Delta^{-1}h'_{\mu\nu},
\end{eqnarray}
and
\begin{eqnarray}
f_{\mu}^{\alpha}N_{(0)}^{\mu}&=&N'^{\alpha},
\end{eqnarray}
where
$h'_{\mu\nu}
=f_{\mu}^{\alpha}\delta_{\alpha\beta}f_{\nu}^{\beta}$.
We can then read off the corresponding vierbein
\begin{eqnarray}
E_M^{~~A}&=&\left(
\begin{array}{ccc}
N'&N'^{\alpha}&N'^3\\
0&f_{\mu}^{\alpha}&B_{\mu}\\
0&0&1
\end{array}
\right)
\left(
\begin{array}{ccc}
\Delta^{-\frac{1}{2}}&&\\
&\Delta^{-\frac{1}{2}}&\\
&&\Delta^{\frac{1}{2}}
\end{array}
\right)\nonumber\\
&\equiv&\left(
\begin{array}{cc}
{}~~N~~&~~N^a~~\\
0&e_m^a
\end{array}
\right).
\label{vierbein}
\end{eqnarray}
Here the Lorentz frame indices $A,B,\ldots$, $a,b,\ldots$
and $\alpha,\beta,\ldots$ take $\{0,1,2,3\}$, $\{1,2,3\}$
and $\{1,2\}$, respectively.
$N'^3$ is sometimes denoted by $B_t$ in this paper, when
it is more appropriate to be regarded as a part of component
of the Kaluza-Klein vector rather than as an element of the
shift.
We further define
\begin{equation}
f_{m'}^{a'}=\left(
\begin{array}{cc}N'&N'^{\alpha}\\
0&f_{\mu}^{\alpha}\end{array}\right),
\end{equation}
where we use $m'=t,x,y$ and $a'=0,1,2$ as the reduced
three-dimensional spacetime and Lorentz frame indices.
Also we write the inverse as
\begin{eqnarray}
E_A^{~M}&=&
\left(
\begin{array}{ccc}
\Delta^{\frac{1}{2}}&&\\
&\Delta^{\frac{1}{2}}&\\
&&\Delta^{-\frac{1}{2}}
\end{array}
\right)
\left(
\begin{array}{ccc}
N'^{-1}&-N'^{-1}N'^{\mu}&N^{-1}(B_{\alpha}N'^{\alpha}-N'^3)\\
0&f_{\alpha}^{\mu}&-B_{\alpha}\\
0&0&1
\end{array}
\right)\nonumber\\
&\equiv&\left(
\begin{array}{cc}
{}~N^{-1}~&-N^{-1}N^m\\
0&e_a^m
\end{array}
\right),\label{inverse}
\end{eqnarray}
where $B_{\alpha}=f_{\alpha}^{\mu}B_{\mu}$.
The upper-left submatrix of the second factor is $f_{a'}^{m'}$.
For convenience for the calculation we write
$e_m^{~a}$ and $e_a^{~m}$ explicitly
\begin{eqnarray}
e_m^{~a}&=&
\left(\begin{array}{cc}
e_{\mu}^{~\alpha}&e_{\mu}^{~3}\\
e_{z}^{~\alpha}&e_{z}^{~3}
\end{array}\right)
{}~=~
\left(\begin{array}{cc}
f_{\mu}^{~\alpha}&B_{\mu}\\
0&1
\end{array}\right)
\left(\begin{array}{cc}
\Delta^{-\frac{1}{2}}&\\
&\Delta^{\frac{1}{2}}
\end{array}\right)
\nonumber\\
&=&
\left(\begin{array}{cc}
\Delta^{-\frac{1}{2}}f_{\mu}^{~\alpha}&\Delta^{\frac{1}{2}}B_{\mu}\\
0&\Delta^{\frac{1}{2}}
\end{array}\right),
\\
e_a^{~m}&=&
\left(\begin{array}{cc}
e_{\alpha}^{~\mu}&e_{\alpha}^{~z}\\
e_{3}^{~\mu}&e_{3}^{~z}
\end{array}\right)
{}~=~
\left(\begin{array}{cc}
\Delta^{\frac{1}{2}}&\\
&\Delta^{-\frac{1}{2}}
\end{array}\right)
\left(\begin{array}{cc}
f_{\alpha}^{~\mu}&-B_{\alpha}\\
0&1
\end{array}\right)\nonumber\\
&=&
\left(\begin{array}{cc}
\Delta^{\frac{1}{2}}f_{\alpha}^{~\mu}
&-\Delta^{\frac{1}{2}}B_{\alpha}\\
0&\Delta^{-\frac{1}{2}}
\end{array}\right).
\end{eqnarray}

Due to the
assumption that we consider the U(1) bundle, the 1-form
\begin{equation}
\eta\equiv dz+B_{m'}dx^{m'}
(=dz+B_{\mu}dx^{\mu}+N'^3dt)
\end{equation}
is a section of the U(1) bundle defined locally on each
coordinate patch. For illustrative purposes
let $\widetilde{\Sigma}$ be ${\rm S}^2$ and
let $H^{(+)}$ and $H^{(-)}$ be its local patches covering each
hemisphere with the intersection
$H^{(+)}\cap H^{(-)}\sim{\rm S}^1$ at the equator in common.
We parameterize this ${\rm S}^1$ by $\theta\in[0,2\pi]$.
Then $\eta$'s defined on each patch differ by the U(1) gauge
transformation
\begin{equation}
\eta^{(+)}=\eta^{(-)}+d\varphi, \label{gaugetr}
\end{equation}
for some $\varphi$ at the equator, satisfying
$\varphi(\theta =2\pi)-\varphi(\theta =0)=2\pi n$, $n\in
{\rm Z}$. This integer, referred to as first Chern class,
characterizes the U(1) bundle under consideration.
A closed two form $\Phi$
obtained by pulling $d\eta$ back to the two-dimensional base
manifold $\widetilde{\Sigma}$ must satisfy the integrality
condition \cite{Moncrief}
\begin{equation}
\int_{\widetilde{\Sigma}}\Phi=2\pi n,
\end{equation}
or, using the component fields,
\begin{equation}
\int_{\widetilde{\Sigma}}(\partial_xB_y-\partial_yB_x)=2\pi n.
\label{integrality}
\end{equation}
It may be easily noticed that this is nothing but the
quantization condition of Kaluza-Klein monopole
\cite{KKmonopole} (Here this is nothing but
the Dirac monopole; see, e.g. \cite{Dirac}). Indeed,
the integrand of (\ref{integrality}) is just the Kaluza-Klein
``magnetic'' field, whose total flux is determined by the
cohomology class of the transition function
(\ref{gaugetr}) characterizing the bundle.
This is the simplest example of the known fact that the
solitonic solution of the Kaluza-Klein field is classified
by $\pi^1$ (first fundamental group) of the isometry group
generated by the Killing vector fields
\cite{P}.
The number of ``monopole
charge'' controls the topology of spacetime; for example, in the
case of $\widetilde{\Sigma}\sim {\rm S}^2$, spacetime is a direct
product ${\rm S}^2\times {\rm S}^1$ if monopole charge is zero,
and ${\rm S}^3$ if monopole charge is one.

In the rest of this paper, we do not consider this
global applicability
\footnote{The possibility of transition between the two U(1)
bundles with distinct monopole charges is examined in
ref. \cite{Moncrief_CQG}. The appearance of unphysical
singularities after Geroch's transformation is discussed
in refs. \cite{Geroch_singularity}.
}
of Geroch's transformation, but restrict
ourselves to focusing on only local properties of the Ashtekar
connection under the transformation.

\subsection{The Ashtekar connection and the Ernst potential}

In order to clarify Geroch's symmetry in the Ashtekar formulation,
let us first express the Ashtekar connection using the
parameterization (\ref{vierbein}). We follow the notation
\cite{NicolaiMatschull} for the Ashtekar formulation in this paper.
the Ashtekar connection is given by
\begin{eqnarray}
A_{ma}&=&
-\frac{1}{2}\epsilon_{abc}\omega_{mbc}\pm 2i\widehat{p}_{ma}
\nonumber\\
&=&-\frac{1}{4}\epsilon_{abc}(2\Omega_{dbc}-\Omega_{bcd})e_m^d
\pm ie_{mb}\Omega_{0(ab)}\label{Ama},
\end{eqnarray}
where $\Omega_{ABC}=2E_{[A}^ME_{B]}^N\partial_ME_{NC}$. The
symmetrization and the anti-symmetrization are the ones of
strength one. $\Omega_{ABC}$ are explicitly written as
\begin{eqnarray}
\Omega_{\alpha\beta\gamma}
&=&2\Delta f_{[\alpha}^{~~\mu}f_{\beta]}^{~~\nu}
\partial_{\mu}(\Delta^{-\frac{1}{2}}f_{\nu\gamma}),\nonumber\\
\Omega_{\alpha\beta 3}
&=&2\Delta^{\frac{3}{2}}f_{[\alpha}^{~~\mu}f_{\beta]}^{~~\nu}
\partial_{\mu}B_{\nu},\nonumber\\
\Omega_{3\beta\gamma}
&=&0,\nonumber\\
\Omega_{3\beta 3}
&=&-f_{\beta}^{\mu}\partial_{\mu}(\Delta^{\frac{1}{2}}),
\nonumber\\
\Omega_{ab0}
&=&0~~~(a,b=\alpha(=1,2), 3),\nonumber\\
\Omega_{0\beta 0}
&=&-\Delta f_{\beta}^{\mu}\partial_{\mu}(\Delta^{-\frac{1}{2}}N'),
\nonumber\\
\Omega_{030}
&=&0,\nonumber\\
\Omega_{033}
&=&
f_{0}^{m'}\partial_{m'}
(\Delta^{\frac{1}{2}}),
\nonumber\\
\Omega_{03\gamma}
&=&0,\nonumber\\
\Omega_{0\beta 3}
&=&
\Delta^{\frac{3}{2}}f_{\beta}^{\nu}f_{0}^{m'}
(\partial_{m'}B_{\nu}-\partial_{\nu}B_{m'})
\nonumber\\
\Omega_{0\beta\gamma}
&=&
\Delta^{\frac{1}{2}}f_{0}^{m'}f_{\beta}^{\nu}
(\partial_{m'}f_{\nu\gamma}-\partial_{\nu}f_{m'\gamma})
-\delta_{\beta\gamma}f_{0}^{m'}\partial_{m'}(\Delta^{\frac{1}{2}}).
\end{eqnarray}
Plugging these expressions into eq.(\ref{Ama}),
we obtain
\begin{eqnarray}
A_{z3}&=&
\pm i \frac{1}{2}f_{0}^{m'}\partial_{m'}\Delta
+\frac{1}{2}\Delta^2 \epsilon_{3\alpha\beta}
f_{\alpha}^{\mu}f_{\beta}^{\nu}\partial_{\mu}B_{\nu},
\label{A_z3}\\
A_{z\alpha}&=&
\frac{1}{2}\epsilon_{3\alpha\beta}f_{\beta}^{\mu}
\partial_{\mu}\Delta
\pm i\frac{1}{2} \Delta^2 f_{\alpha}^{\nu}f_{0}^{m'}
(\partial_{m'}B_{\nu}-\partial_{\nu}B_{m'}).
\label{A_zalpha}
\end{eqnarray}
To express these formulas in terms of the Ernst potential
${\cal E}_{\pm} \equiv \Delta\pm iB$, we now invoke the duality
relation
\newcounter{tmpequation}
\setcounter{tmpequation}{\theequation}
\setcounter{equation}{\theduality}
\begin{equation}
\Delta^2(\partial_{m'}B_{n'}-\partial_{n'}B_{m'})
=f_{m'}^{a'}f_{n'}^{b'}f_{p'}^{c'}\epsilon_{a'b'c'}
\partial^{p'}B.
\end{equation}
\setcounter{equation}{\thetmpequation}
Use of this equation leads to the following simple result
\begin{eqnarray}
A_{z3}&=&
\pm i\frac{1}{2}f_{0}^{m'}
\partial_{m'}{\cal E}_{\pm},\\
A_{z\alpha}&=&
\frac{1}{2}\epsilon_{3\alpha\beta}f_{\beta}^{\mu}
\partial_{\mu}{\cal E}_{\pm},
\end{eqnarray}
where $\pm$ depends on the choice of the sign in (\ref{Ama}).
The third column of the Ashtekar connection is therefore
nothing but (essentially) the gradient of the Ernst potential.
What may be a remarkable thing is that the Ashtekar connection is
readily complexified and partially dualized (in the sense that
the second term of the Ashtekar connection has a factor
$\epsilon_{abc}$) to give directly (the
gradient of) the complex Ernst potential. This is the first
observation that shows the close relationship between the Ashtekar
connection and the Ernst potential.  Consequently $A_{z3}$ and
$A_{z\alpha}$ transform as
\begin{eqnarray}
\pm i{\cal E_{\pm}}&\rightarrow&\frac{\pm ia{\cal E_{\pm}}+b}
{\pm ic{\cal E_{\pm}}+d},\nonumber\\
A_{z3}&\rightarrow&\frac{1}{(\pm ic{\cal E}_{\pm} +d)^2}A_{z3},
\nonumber\\
A_{z\alpha}&\rightarrow
&\frac{1}{(\pm ic{\cal E}_{\pm} +d)^2}A_{z\alpha},
\label{Ehlers'}
\end{eqnarray}
where $a,b,c,d\in{\rm R}$ and $ad-bc=1$.

Remaining components of the Ashtekar connection
$A_{\mu\alpha}$ and $A_{\mu 3}$ are linear combinations
of $A^{\mu}_{~\alpha}$ and $A^{\mu}_{~ 3}$, which are calculated
to be
\begin{eqnarray}
A^{\mu}_{~\alpha}&=&
\mp i \frac{1}{2}f_{\alpha}^{\mu}f_{0}^{m'}\partial_{m'}\Delta
-\frac{1}{2}\Delta^2 \epsilon_{3\gamma\beta}f_{\alpha}^{\mu}
f_{\gamma}^{\lambda}f_{\beta}^{\nu}\partial_{\lambda}B_{\nu}
\pm i\Delta\omega^{(f)\mu}_{~~~~\alpha 0},
\label{A^mu_alpha}
\\
A^{\mu}_{~3}&=&
\frac{1}{2}\epsilon_{3\alpha\beta}
f_{\alpha}^{\mu}f_{\beta}^{\nu}
\partial_{\nu}\Delta
\pm i\frac{1}{2} \Delta^2
f_{\alpha}^{\mu}f_{\alpha}^{\nu}f_{0}^{m'}
(\partial_{m'}B_{\nu}-\partial_{\nu}B_{m'})
-\Delta\omega^{(f)\mu}_{~~~~12},
\label{A^mu_3}
\end{eqnarray}
where
$\omega^{(f)m'}_{~~~~b'c'}$
is the spin-connection with respect to $f_{m'}^{a'}$
\begin{eqnarray}
\omega^{(f)m'}_{~~~~b'c'}
&=&
\frac{1}{2}(\Omega^{(f)}_{a'b'c'}
-\Omega^{(f)}_{b'c'a'}
+\Omega^{(f)}_{c'a'b'})f^{m'a'},\\
\Omega^{(f)}_{a'b'c'}
&=&
2f_{[a'}^{m'}f_{b']}^{n'}\partial_{m'}f_{n'c'}
\end{eqnarray}
and $B^{\mu}=f_{\alpha}^{\mu}B_{\alpha}$.
The index $\mu$ of the Ashtekar connection is raised by
\begin{eqnarray}
g^{mn}&=&e_a^me_b^n\delta^{ab}\nonumber\\
&=&\left(
\begin{array}{cc}
\Delta f^{\mu}_{~\alpha}f^{\alpha\nu}&-\Delta B^{\mu}\\
-\Delta B^{\nu}&\Delta B_{\alpha}B^{\alpha}+\Delta^{-1}
\end{array}
\right),
\end{eqnarray}
which is the inverse of the submatrix of the original
metric
\begin{eqnarray}
g_{mn}&=&e_m^ae_n^b\delta_{ab}\nonumber\\
&=&\left(
\begin{array}{cc}
\Delta^{-1}f_{\mu}^{~\alpha}f_{\alpha\nu}
+\Delta B_{\mu}B_{\nu}
&\Delta B_{\mu}\\
\Delta B_{\nu}&\Delta
\end{array}
\right).
\end{eqnarray}
Making use of eq.(\ref{duality}) in
(\ref{A^mu_alpha}) and (\ref{A^mu_3}), we find
\begin{eqnarray}
A^{\mu}_{~\alpha}&=&
\pm i\left[
-\frac{1}{2}f_{\alpha}^{\mu}f_{0}^{m'}
\partial_{m'}{\cal E}_{\pm}
+\Delta\omega^{(f)\mu}_{~~~~\alpha 0}
\right],
\\
A^{\mu}_{~3}&=&\frac{1}{2}\epsilon_{3\alpha\beta}
\left[
f_{\alpha}^{\mu}f_{\beta}^{\nu}
\partial_{\nu}{\cal E}_{\pm}
-\Delta \omega^{(f)\mu}_{~~~~\alpha\beta}
\right].
\label{Amu3}
\end{eqnarray}
Thus $A^{\mu}_{~\alpha}$ and $A^{\mu}_{~3}$ can be also expressed
in compact forms in terms of the Ernst potential.
$A_{\mu\alpha}$ and $A_{\mu 3}$ are,
on the other hand, related to them by
\begin{eqnarray}
A_{\lambda 3}&=&
\Delta^{-1}f_{\lambda}^{~\beta}f_{\mu\beta}A^{\mu}_{~3}
+B_{\lambda}A_{z3},\nonumber\\
A_{\lambda \alpha}&=&
\Delta^{-1}f_{\lambda}^{~\beta}f_{\mu\beta}A^{\mu}_{~\alpha}
+B_{\lambda}A_{z\alpha}. \label{formula1}
\end{eqnarray}
Their explicit dependence on $B_{\mu}$ shows that they transform
non-locally under (\ref{Ehlers'}). We will see, however, that
such $B_{\mu}$ dependence will be drop in all the fist class
constraints.

Comparing (\ref{A_zalpha}) (\ref{A_z3})
and (\ref{A^mu_alpha}) (\ref{A^mu_3}),
we find that the following relations hold {\em without using
the duality relation}
\begin{eqnarray}
A^{\mu}_{~\alpha}&=&
-f^{\mu}_{\alpha}A_{z3}\pm i\Delta\omega^{(f)\mu}_{~~~~\alpha 0},
\nonumber\\
A^{\mu}_{~3}&=&
f^{\mu}_{\alpha}A_{z\alpha}-\Delta\omega^{(f)\mu}_{~~~~12}.
\label{formula2}
\end{eqnarray}
These relations will turn out to be useful in a moment.

Before concluding this subsection, a comment is in order.
We have succeeded to relate the Ashtekar connections to the Ernst
potential with the help of the duality relation (\ref{duality}),
which is originally a solution of the field equation of $B_{m'}$
in the Lagrangian formalism.  Therefore, in the framework of
Hamiltonian formalism, we have to make the origin of this equation
clear. As shown in ref.\cite{Moncrief}, {\em a part} of the
relations can be obtained by solving the diffeomorphism
constraint of $z$-coordinate, while $r$ defined by
\begin{eqnarray}
r&\equiv&f f_{\alpha}^{\mu}f_{\beta}^{\nu}
\epsilon_{3\alpha\beta}\partial_{\mu}B_{\nu}\label{r},\\
f&\equiv& {\rm det}f_{\mu}^{\alpha}
\end{eqnarray}
is treated as an
independent canonical variable, whose transformation rule should
be imposed on itself. We will elaborate further on this point in
the next subsection.

\subsection{Ehlers' SL(2, R) symmetry in the Ashtekar
formulation}
We will now show Ehlers' SL(2,R) symmetry of this model within
the framework of Hamiltonian formalism. In the Hamiltonian
formalism in general, we have to clarify the following two points
to prove a symmetry of the system. First, we must show, of course,
the invariance of the Hamiltonian (which is zero in the context of
gravity) and the constraints under the transformation.  Second, we
have to make sure that the new variables resulting from the action
on the canonical variables again satisfy the canonical Poisson
algebra. We do not directly take these steps, but rather we will
do it alternatively in the following way:
we first look for canonical pairs which
transform in a simple way under the transformation, show the
invariance of their Liouville form, and then we rewrite the
constraints in terms of these new canonical variables and see the
invariance of the constraints. The advantage of this approach is
that the second point we mentioned above is automatically guaranteed
by the invariance of the Liouville form, and also, of course, that
the invariance of the constraints is expected to be seen more neatly
in terms of these special canonical variables than the original
ones, which transform in a non-trivial manner.
Ehlers' SL(2,R) invariance is already shown in
ref.\cite{Moncrief} in this way in the ADM formalism. In this paper,
however, we will use the Ashtekar formulation here, in which the
Ernst potential and its complex conjugate will naturally arise as
such canonical variables on which Ehlers' SL(2,R) acts as a linear
fractional transformation.

We start from the parameterization of the inverse vierbein
(\ref{inverse}). The densitized (inverse) dreibein reads
\begin{equation}
\widetilde
{e}_a^{~m}=
f\left(\begin{array}{cc}
f_{\alpha}^{~\mu}&-B_{\alpha}\\
0&\Delta^{-1}
\end{array}\right).
\end{equation}
In the Ashtekar formulation the first class constraints are
given by \cite{Ashtekar}
\begin{eqnarray}
{\cal H}&\equiv&\epsilon^{abc}\widetilde{e}_a^{~m}
\widetilde{e}_b^{~n}F_{mnc},\nonumber\\
{\cal H}_n&\equiv&\widetilde{e}_a^{~m}F_{mn}^{~~a},\nonumber\\
{\cal G}_a&\equiv&D_m\widetilde{e}_a^{~m}.
\end{eqnarray}
${\cal G}_a$ is the Lorentz constraint,
while ${\cal H}$ and ${\cal H}_n$ are the Hamiltonian and
diffeomorphism
constraint modulo the Lorentz constraint. The diffeomorphism is
actually generated by the following linear combination
\begin{equation}
{\cal C}_n=A_{na}{\cal G}_a+{\cal H}_n.
\end{equation}

Since we have taken
$\widetilde{e}_3^{~\mu}=0$, three of the two Lorentz constraints
\begin{equation}
{\cal G}_{\alpha}\equiv D_m\widetilde{e}_{\alpha}^m=0
\label{G_alpha}
\end{equation}
become second class, while other constraints still remain first
class. Solving (\ref{G_alpha}), $A_{\mu 3}$ are written in terms
of other canonical variables
\begin{equation}
\widetilde{e}_{\alpha}^{\mu}A_{\mu 3}
=
-\epsilon_{3\alpha\beta}\partial_{\nu}\widetilde{e}_{\beta}^{\nu}
+A_{z\alpha}\widetilde{e}_{3}^{z}
-A_{z3}\widetilde{e}_{\alpha}^{z}. \label{solvedAmu3}
\end{equation}
We may use this equality in the strong sense to eliminate
$A_{\mu 3}$ in what follows.

        We next consider ${\cal C}_z$. In our case it is reduced
to
\begin{eqnarray}
{\cal C}_z&=&
\widetilde{e}_{\alpha}^{~\mu}
\partial_{\mu}A_{z\alpha}
-\epsilon_{\alpha\beta 3}A_{z\beta}
(\widetilde{e}_{\alpha}^{~\mu}A_{\mu 3}
+\widetilde{e}_{\alpha}^{~z}A_{z 3})\nonumber\\
&=&\partial_{\mu}(\widetilde{e}_{\alpha}^{~\mu}A_{z \alpha}),
\label{Cz}
\end{eqnarray}
where we used (\ref{solvedAmu3})
and the assumption that any field does not depend on
$z$-coordinate. Now let us assume here that we have imposed
any gauge-fixing condition on the $z$-coordinate diffeomorphism
degrees of freedom, so that the constraint ${\cal C}_z$
is already of second class and can be imposed strongly.
For example, one may take
\begin{equation}
N'^z=0
\end{equation}
or equivalently,
\begin{equation}
N^{\alpha}\widetilde{e}_{\alpha}^{~z}
+N^{3}\widetilde{e}_{3}^{~z}
=0
\end{equation}
as a gauge-fixing condition. Alternatively, one may choose
\begin{equation}
\widetilde{e}_{1}^{~z}=0
\end{equation}
as was done in ref.\cite{HS}\footnote{Note that
$\widetilde{e}_{1}^{~y}$
was also set to zero to fix ${\cal C}_y$ in ref.\cite{HS} since
they considered a two-Killing reduced model. We also take this
gauge-choice in the next section.}.
In any case, one may easily verify that the constraint
${\cal C}_z$ together with either of these conditions actually
become second class. We may then solve the equation
\begin{equation}
{\cal C}_z=0 \label{Cz=0}
\end{equation}
explicitly. Due to Stokes' theorem,
(\ref{Cz=0}) with (\ref{Cz})
means that there exists some $\phi$ such that
$\widetilde{e}_{\alpha}^{~\mu}A_{z \alpha}$ can be
written locally
\begin{equation}
\widetilde{e}_{\alpha}^{~\mu}A_{z \alpha}
=ff_{\alpha}^{\mu}f_{\beta}^{\nu}\epsilon_{3\alpha\beta}
\partial_{\nu}\phi.
\label{phi}
\end{equation}

In fact, $\phi$ is nothing but (locally) a half of the Ernst
potential: $\phi={\cal E}_{\pm}/2$ ($\pm$ depends on the choice of
the sign in the definition of the Ashtekar connection (\ref{Ama})),
as can be checked by substituting (\ref{A_zalpha}) into (\ref{phi}).


In the ADM
formalism, the diffeomorphism constraint of $z$-coordinate
amounts to a requirement that the $B_{\mu}$ field should be
divergence free \cite{Moncrief}. This can be similarly solved to
ensure that the $B_{\mu}$ can be written as a Hodge dual
of the gradient of $B$\footnote{$B$ is denoted by
$\widetilde{\psi}$ in ref.\cite{BM}, and $\omega$ in
ref.\cite{Moncrief}.},
which is an imaginary part of the Ernst potential. In our case,
we have used complex
the Ashtekar connection to write down ${\cal C}_z$ to find
$\widetilde{e}_{\alpha}^{~\mu}A_{z \alpha}$ be divergence free,
being led directly to the complex Ernst potential.
Hence, in that sense, both the Ashtekar connection and the Ernst
potential are complexified ``in the same way''.

The equation (\ref{phi}) reproduces only a part of the duality
relations (\ref{duality}). To see this, let us write
$\widetilde{e}_{\alpha}^{~\mu}A_{z \alpha}$, using
(\ref{A_zalpha}), explicitly in terms of the components
of dreibein
\begin{equation}
\widetilde{e}_{\alpha}^{~\mu}A_{z \alpha}
=
ff_{\alpha}^{\mu}
\left(
\frac{1}{2}\epsilon_{3\alpha\beta}f_{\beta}^{\mu}
\partial_{\mu}\Delta
\pm i\frac{1}{2} \Delta^2 f_{\alpha}^{\nu}f_{0}^{m'}
(\partial_{m'}B_{\nu}-\partial_{\nu}B_{m'})
\right).
\end{equation}
The first term is already in the form of
a rotation. Hence (\ref{phi}) forces the second term to be
$
ff_{\alpha}^{\mu}f_{\beta}^{\nu}\epsilon_{3\alpha\beta}
\partial_{\nu}B
$
for some $B$. This may be obtained by setting $a'=0$
in the equation below equivalent to (\ref{duality})
\begin{equation}
\Delta^2f_{a'}^{m'}(\partial_{m'}B_{\mu}-\partial_{\mu}B_{m'})
=f_{\mu}^{\beta}f_{p'}^{c'}\epsilon_{a'\beta c'}
\partial^{p'}B. \label{duality2}
\end{equation}
On the other hand, the equation for the $xy$-space rotation of
$B_{\mu}$ can not be derived by (\ref{phi}), but rather such a
field is contained in the degrees of freedom of canonical
momenta, whose transformation rule should be independently
imposed.

Let us now look for ``good'' canonical pairs for Ehlers'
SL(2,R). Since we would like to take
$\phi(={\cal E}_{\pm}/2)$ as a canonical
variable, we write the Liouville form in terms of
$\widetilde{e}_{\alpha}^{~\mu}A_{z \alpha}$ as a first step
\begin{eqnarray}
&&\dot{A_{\mu\alpha}}\widetilde{e}_{\alpha}^{\mu}
+\dot{A_{z\alpha}}\widetilde{e}_{\alpha}^{z}
+\dot{A_{z3}}\widetilde{e}_{3}^{z}\nonumber\\
&=&-\dot{ (\widetilde{e}_{\alpha}^{~\mu}A_{z \alpha}) }B_{\mu}
+\dot{\widetilde{e}_{\alpha}^{\mu}}
(A_{z\alpha}B_{\mu}-A_{\mu\alpha})
+\dot{A_{z3}}\widetilde{e}_{3}^{z}\nonumber\\
&=&-\dot{\phi}r
+\dot{\overbrace{(A_{\mu\alpha}-A_{z\alpha}B_{\mu})}}
{}~\widetilde{e}_{\alpha}^{\mu}
+\dot{A_{z3}}\widetilde{e}_{3}^{z},
\end{eqnarray}
where the equalities hold up to total derivatives.
$r$ is defined by (\ref{r}) in the previous section.
Using the formulas (\ref{formula1})(\ref{formula2}),
this is further rewritten in the following form
\begin{equation}
=-\dot{\phi}r
+\dot{(fA_{z3})}f^{-1}
\widetilde{e}_{3}^{z}
\pm i\dot{\omega^{(f)}_{\mu\alpha 0}}\widetilde{e}_{\alpha}^{\mu},
\label{Liouvilletmp}
\end{equation}
with $\pm$ depending on the sign in the definition of the Ashtekar
connection.
Finally the relation
\begin{equation}
A_{z3}-\Delta^2f^{-1}r=-\overline{A_{z3}} \label{Abar},
\end{equation}
allows us to rewrite (\ref{Liouvilletmp}) as
\begin{equation}
=f\Delta^{-2}(\dot{\overline{\phi}}A_{z3}
-\dot{\phi}\overline{A_{z3}})
\pm i\dot{\omega^{(f)}_{\mu\alpha 0}}
\widetilde{e}_{\alpha}^{\mu}.
\label{LformEhler}
\end{equation}
The relation (\ref{Abar}) can be immediately shown from
(\ref{A_z3}).  Since the first two terms of (\ref{LformEhler}) give
their imaginary part, the total Liouville form is purely imaginary.
The new canonical variables are
thus $(\overline{\phi},f\Delta^{-2}A_{z3})$,
$(\phi,-f\Delta^{-2}\overline{A_{z3}})$ and
$(\pm i\omega^{(f)}_{\mu\alpha 0},
\widetilde{e}_{\alpha}^{\mu})$.
The Liouville form (\ref{LformEhler}) is invariant if these
canonical variables transform as follows
\begin{eqnarray}(\overline{\phi},f\Delta^{-2}A_{z3})
&\rightarrow&
\left(
\frac{1}{i}\cdot
\frac{ia\overline{\phi}+b}
{ic\overline{\phi}+d},~
(ic\overline{\phi}+d
)^2\cdot f\Delta^{-2}A_{z3}
\right),\nonumber\\
(\phi,-f\Delta^{-2}\overline{A_{z3}})
&\rightarrow&
\left(
-\frac{1}{i}\cdot
\frac{-ia\phi+b}
{-ic\phi+d},~
(-ic\phi+d
)^2\cdot (-f\Delta^{-2}\overline{A_{z3}})
\right),\nonumber\\
(\pm i\omega^{(f)}_{\mu\alpha 0},
\widetilde{e}_{\alpha}^{\mu})
&\rightarrow&
(\pm i\omega^{(f)}_{\mu\alpha 0},
\widetilde{e}_{\alpha}^{\mu}).
\label{Ehlercanvar}
\end{eqnarray}
This is Ehlers' SL(2,R) transformation in the Hamiltonian
formalism.
What we have seen here is that we could take the Ernst
potential and its complex conjugate as canonical variables
(although we have used its half $\phi$ actually to avoid the
appearance of the factor 2 everywhere), and then their canonical
conjugate turned out to be proportional to the $A_{z3}$ and the
$\overline{A_{z3}}$. The complex conjugate field has naturally
appeared to give manifestly pure imaginary Liouville form and,
as we see below, the constraints. These phenomena also reflect a
nice structure of Ehlers' symmetry in the Ashtekar
formulation. This transformation is generically non-local
with respect to the original Ashtekar variables.

We will now write the first class constraints in manifestly
invariant forms in terms of the canonical variables above.
The Hamiltonian constraint is reduced to
\begin{equation}
{\cal H}=\Delta^{-1}f^2\left[
2\epsilon_{3\alpha\beta}e_3^ze_{\alpha}^{\mu}F_{z\mu\beta}
+2\epsilon_{3\alpha\beta}e_{\alpha}^ze_{\beta}^{\mu}F_{z\mu3}
+\epsilon_{3\alpha\beta}
e_{\alpha}^{\mu}e_{\beta}^{\nu}F_{\mu\nu 3}
\right],
\label{HaEhlertmp}
\end{equation}
where
\begin{eqnarray}
F_{z\mu\beta}&=&-\partial_{\mu}A_{z\beta}
+\epsilon_{\beta\gamma 3}(A_{z\gamma}A_{\mu 3}
-A_{z3}A_{\mu\gamma}),\nonumber\\
F_{z\mu 3}&=&-\partial_{\mu}A_{z3}
+\epsilon_{3\alpha\beta}A_{z\alpha}A_{\mu\beta},
\nonumber\\
F_{z\mu 3}&=&\partial_{\mu}A_{\nu 3}-\partial_{\nu}A_{\mu3}
+\epsilon_{3\alpha\beta}A_{\mu\alpha}A_{\nu\beta}.
\label{FEhler}
\end{eqnarray}
There appears no $F_{\mu\nu\alpha}$. Substituting
(\ref{formula1})(\ref{formula2}) into
(\ref{HaEhlertmp})(\ref{FEhler}), we find that all the explicit
dependence on $B_{\mu}$ cancel, obtaining
\begin{eqnarray}
{\cal H}&=&
-f^2\left[
2\Delta^{-2}(-A_{z\alpha}\overline{A_{z\alpha}}
-A_{z3}\overline{A_{z3}})
+\epsilon_{3\alpha\beta}f_{\alpha}^{\mu}f_{\beta}^{\nu}
(\partial_{\mu}\omega_{\nu 12}^{(f)}
-\partial_{\nu}\omega_{\mu 12}^{(f)})
+\epsilon_{3\alpha\beta}\epsilon_ {3\gamma\delta}
\omega_{\alpha\gamma 0}^{(f)}
\omega_{\beta\delta 0}^{(f)}
\right]\nonumber\\
&=&\frac{1}{2}(\phi+\overline{\phi})^{-2}
\widetilde{e}_{\alpha}^{\mu}
\widetilde{e}_{\alpha}^{\nu}
\partial_{\mu}\phi
\partial_{\mu}\overline{\phi}
+\frac{1}{2}(\phi+\overline{\phi})^2
(f\Delta^{-2}A_{z3})(f\Delta^{-2}\overline{A_{z3}})
\nonumber\\&&
-2\epsilon_{3\alpha\beta}
\widetilde{e}_{\alpha}^{\mu}
\widetilde{e}_{\alpha}^{\nu}
\partial_{\mu}\omega_{\nu 12}^{(f)}
-\epsilon_{3\alpha\beta}\epsilon_ {3\gamma\delta}
\widetilde{e}_{\alpha}^{\mu}\omega_{\mu\gamma 0}^{(f)}
\widetilde{e}_{\beta}^{\nu}\omega_{\nu\delta 0}^{(f)},
\label{HaEhler}
\end{eqnarray}
where in the first line we have used
\begin{equation}
A_{z\alpha}-\epsilon_{3\alpha\beta}f_{\beta}^{\mu}\partial_{\mu}
\Delta=-\overline{A_{z\alpha}}.
\end{equation}
The last two terms of (\ref{HaEhler}) are trivially invariant
(Note that $\omega^{(f)}_{\nu 12}$ is a
function of $\widetilde{e}_{\alpha}^{\mu}$.).
The first two are also invariant since
\begin{eqnarray}
\phi+\overline{\phi}
&\rightarrow& \frac{\phi+\overline{\phi}}{|c\phi+id|^2},\nonumber\\
\partial_{\mu}\overline{\phi}
&\rightarrow& \frac{\partial_{\mu}\overline{\phi}}
{(ic\overline{\phi}+d)^2},\nonumber\\
\partial_{\mu}\phi
&\rightarrow& \frac{\partial_{\mu}\phi}
{(ic\phi-d)^2}.
\end{eqnarray}
Thus the expression (\ref{HaEhler}) is invariant under Ehlers'
SL(2,R)
transformation (\ref{Ehlercanvar}). (\ref{HaEhler})
is independent of the sign choice of the Ashtekar connection.

The diffeomorphism constraints $\cal{C}_{\mu}$ can be also
made manifestly symmetric
\begin{equation}
{\cal C}_{\mu}=\pm i \left\{
\partial_{\nu}(\widetilde{e}_{\alpha}^{\nu}
\omega^{(f)}_{\mu\alpha 0})
-\widetilde{e}_{\alpha}^{\nu}
\partial_{\mu}\omega^{(f)}_{\nu\alpha 0}
\right\}
-(\partial_{\mu}\overline{\phi}\cdot
f\Delta^{-2}A_{z3}
-\partial_{\mu}\phi\cdot
f\Delta^{-2}\overline{A_{z3}}).
\end{equation}

Finally the Lorentz constraint ${\cal G}_3$ is simply written as
\begin{equation}
{\cal G}_3=\pm i\epsilon_{3\alpha\beta}
\widetilde{e}_{\beta}^{\mu}\omega^{(f)}_{\mu\alpha 0}.
\end{equation}
This completes the proof of Ehlers' SL(2,R) symmetry in the
Ashtekar formulation.

\subsection{Conserved charges}
        It is now easy to calculate the conserved charges
associated with the invariance under (\ref{Ehlercanvar}).
Corresponding to the Chevalley generators of sl(2,R)
\begin{equation}
h\equiv\left(
\begin{array}{cc}
1&\phantom{0}\\ \phantom{0}&-1
\end{array}
\right),~
e\equiv\left(
\begin{array}{cc}
\phantom{0}&\phantom{0}1\\\phantom{0}&\phantom{0}
\end{array}
\right),~
f\equiv\left(
\begin{array}{cc}
\phantom{0}&\phantom{0}\\1&\phantom{-0}
\end{array}
\right),\label{SL(2,R)}
\end{equation}
we will write the infinitesimal
Ehlers' SL(2,R) action
on the canonical variables as
$h^{(E)}$, $e^{(E)}$, $f^{(E)}$,
which we may read off from (\ref{Ehlercanvar}).
We have
\begin{eqnarray}
&&h^{(E)}(\overline{\phi})
=
-2\overline{\phi},~~
e^{(E)}(\overline{\phi})
=\frac{1}{i},~~
f^{(E)}(\overline{\phi})
=\frac{1}{i}
\overline{\phi}^2, \nonumber\\
&&h^{(E)}(\phi)
=
-2\phi,~~
e^{(E)}(\phi)
=-\frac{1}{i},~~
f^{(E)}(\phi)
=-\frac{1}{i}
\phi^2,
\end{eqnarray}
while
\begin{equation}
h^{(E)}(\pm i\omega^{(f)}_{\mu\alpha 0})
=e^{(E)}(\pm i\omega^{(f)}_{\mu\alpha 0})
=f^{(E)}(\pm i\omega^{(f)}_{\mu\alpha 0})
=0.
\end{equation}
Setting $f\Delta^{-2}A_{z3}\equiv p_{\overline{\phi}}$,
$-f\Delta^{-2}\overline{A_{z3}}\equiv p_{\phi}$,
the conserved charges are given by
\begin{eqnarray}
K_{h^{(E)}}&\equiv&\int -2\left(p_{\overline{\phi}} \overline{\phi}
+ p_{\phi} \phi\right),\nonumber\\
K_{e^{(E)}}&\equiv&\int \frac{1}{i}
\left(p_{\overline{\phi}} -p_{\phi}\right),\nonumber\\
K_{f^{(E)}}
&\equiv&\int
\frac{1}{i}
\left(p_{\overline{\phi}}\overline{\phi}^2-p_{\phi}\phi^2\right).
\label{Ehlerconservedcharges}
\end{eqnarray}
It is an easy exercise to check that their Poisson bracket
algebra satisfy the commutation relation of sl(2,R).
We may also verify that these conserved charges
coincide the ones obtained in ref.\cite{Moncrief} up to
trivial constant factors. We will see in the next section
that the conserved charges associated with Matzner-Misner's
SL(2,R) can be expressed precisely in the same form as above.

\section{Reduction from four to two dimensions}
\label{4to2}
\subsection{Invariance under Matzner-Misner's SL(2,R)}
To reduce the spacetime dimensions to two,  we
introduce another Killing vector field on $\widetilde{\Sigma}$.
Since we do not discuss the global applicability of the Geroch
group, we simply assume that we may take a global
coordinate $y$ such that $\frac{\partial}{\partial y}$
is a Killing vector field. $\widetilde{\Sigma}$ is not
necessarily be compact.
We adopt the same gauge-fixing condition as was set in
ref.\cite{HS} to fix the diffeomorphism degrees of freedom
of $y$- and $z$-coordinates
\begin{equation}
\widetilde{e}_1^y=\widetilde{e}_1^z=0.
\end{equation}
These conditions allow us to solve
${\cal C}_y={\cal C}_z=0$,
which are now of second class. The solution is
found to be \cite{HS}
\begin{equation}
A_{y1}=A_{z1}=0. \label{A_y,z1=0}
\end{equation}
They simply determine the Lagrange parameters $N^y$ and $N^z$
so that $\Omega_{10\bar{c}}=0$. This means that
$\partial_xN^y=\partial_xN^z=0$.
In ref.\cite{HS} the gauge-fixing condition for the Lorentz
constraints ${\cal G}_2$, ${\cal G}_3$ are also set by requiring
\begin{equation}
\widetilde{e}_2^x=\widetilde{e}_3^x=0. \label{e_2,3^x=0}
\end{equation}
We have, however, already set
$\widetilde{e}_2^x=\widetilde{e}_3^y=0$ to distinguish the
coordinate of the U(1) fiber. Consequently there remain
no degrees of freedom of Lorentz rotations. The solution
for the Lorentz constrains ${\cal G}_2={\cal G}_3=0$ is \cite{HS}
\begin{equation}
A_{x2}=A_{x3}=0.
\end{equation}
It turns out that these equations arise no further restriction
on the Lagrange parameters than the ones arising from
(\ref{A_y,z1=0}). To summarize, the non-vanishing components of
the densitized inverse dreibein are
\begin{equation}
\widetilde{e}_a^m=\left(
\begin{array}{ccc}
\widetilde{e}_1^x&&\\
&\widetilde{e}_2^y&\widetilde{e}_2^z\\
&&\widetilde{e}_3^z
\end{array}
\right)
=
\left(
\begin{array}{ccc}
\rho&&\\
&f_x^1&-f_x^1\psi\\
&&f_x^1\frac{\rho}{\Delta}
\end{array}
\right),\label{MMparametrization}
\end{equation}
where we also wrote their parameterization suitable for the
description of Matzner-Misner's SL(2,R) symmetry. The vanishing
Ashtekar connections are
\begin{equation}
A_{x2}=A_{x3}=A_{y1}=A_{z1}=0.
\end{equation}
The independent canonical pairs are thus
$(\widetilde{e}_1^x, A_{x1})$,
$(\widetilde{e}_2^y, A_{y2})$,
$(\widetilde{e}_2^z, A_{x1})$ and $(\widetilde{e}_3^z, A_{z3})$.
Finally, $A_{y3}$ is non-vanishing, but is not independent,
obeying
\begin{equation}
\widetilde{e}_2^y A_{y3}
=\partial_x\widetilde{e}_1^x
+\widetilde{e}_3^zA_{z2}-\widetilde{e}_2^zA_{z3}.
\label{solvedA_y3}
\end{equation}

In the similar way we have done in the previous section, let us
first write the Ashtekar connections in terms of the
parameterization
(\ref{MMparametrization}).
The coefficients of anholonomy are in this case
\begin{eqnarray}
\Omega_{1\bar{b}\bar{c}}
&=&
e_1^x\bar{e}_{\bar{b}}^{\bar{n}}
\partial_x\bar{e}_{\bar{n}\bar{c}},\nonumber\\
\Omega_{0\bar{b}\bar{c}}
&=&
E_0^{\tilde{m}}\bar{e}_{\bar{b}}^{\bar{n}}
\partial_{\tilde{m}}\bar{e}_{\bar{n}\bar{c}},\nonumber\\
\Omega_{10\bar{c}}
&=&
e_1^x(E_0^t\partial_xE_{t\bar{c}}
+E_0^{\bar{n}}\partial_x\bar{e}_{\bar{n}\bar{c}})=0,\nonumber\\
\Omega_{101}
&=&
e_1^xE_0^{\tilde{n}}
(\partial_xE_{\tilde{n}1}
-\partial_{\tilde{n}}E_{x1}),\nonumber\\
\Omega_{100}
&=&
-e_1^xN^{-1}\partial_x N,\nonumber\\
\mbox{\rm otherwise}&=&0,
\end{eqnarray}
where the dreibein is now
\footnote{We use the notation $\bar{e}_{{\bar{m}}}^{\bar{a}}$
for the lower right two by two block of the dreibein,
following ref.\cite{BM}. This is, of course, no complex conjugate
of anything.}
\begin{eqnarray}
e_m^a&=&\left(
\begin{array}{cc}
e_x^1&\\
&\bar{e}_{{\bar{m}}}^{\bar{a}}\\
\end{array}
\right)\nonumber\\
&=&
\left(
\begin{array}{ccc}
f_x^1&&\\
&\rho&\psi\\
&&1
\end{array}
\right)
\left(
\begin{array}{ccc}
\Delta^{-\frac{1}{2}}&&\\
&\Delta^{-\frac{1}{2}}&\\
&&\Delta^{\frac{1}{2}}
\end{array}
\right).
\end{eqnarray}
We have set $f_y^2\equiv\rho$, $B_y\equiv\psi$ for simpler
notations (but will leave $f_x^1$ as it is, to remember that
it belongs to the ``uncompactified'' sector, on which the center
of the Geroch group acts as a scale transformation, as we shall
see below). The Ashtekar connections are calculated to be
\begin{eqnarray}
A_{z3}&=&
\frac{\alpha}{2}f_0^{\widetilde{m}}\partial_{\widetilde{m}}\Delta
+\frac{1}{2}\Delta^2\rho^{-1}f_1^x\partial_x \psi,\nonumber\\
A_{z2}&=&-\frac{1}{2}f_1^x\partial_x\Delta
+\frac{\alpha}{2}\Delta^2\rho^{-1}f_0^{\widetilde{m}}
\partial_{\widetilde{m}} \psi,\nonumber\\
A_{y3}&=&\Delta^{-1}\rho A_{z2}+\psi A_{z3}+f_1^x\partial_x\rho,
\nonumber\\
A_{y2}&=&-\Delta^{-1}\rho A_{z3}
+\psi A_{z2}+\alpha f_0^{\widetilde{m}}\partial_{\widetilde{m}}\rho,
\nonumber\\
A_{x1}&=&-\frac{1}{2}\frac{\Delta}{\rho}\partial_x{\psi}
+\frac{\alpha\sqrt{\Delta}}{N'}
(\partial_t(\frac{f_x^1}{\sqrt{\Delta}})
-\partial_x(\frac{N'_1}{\sqrt{\Delta}})
),\label{AshtekarMM}
\end{eqnarray}
where, from the definition (\ref{MMparametrization}),
\begin{eqnarray}
\rho&=&\widetilde{e}_1^x,\nonumber\\
f_x^1&=&\widetilde{e}_2^y,\nonumber\\
\psi&=&-\frac{\widetilde{e}_2^z}{\widetilde{e}_2^y},\nonumber\\
\Delta&=&
\frac{\widetilde{e}_1^x\widetilde{e}_2^y}{\widetilde{e}_3^z},
\end{eqnarray}
and $\alpha$ stands for $\pm i$ in the definition of the Ashtekar
connection (\ref{Ama}). The expression (\ref{AshtekarMM}) is not
very illuminating. However, let us consider the following
combinations
\begin{eqnarray}
(\pm\pm)&\equiv&\widetilde{e}_{\pm}^{{\bar{m}}}A_{{\bar{m}}\pm}
\nonumber\\
&=&
\pm i f_x^1\rho(\alpha f_0^{\widetilde{m}}\pm if_1^{\widetilde{m}})
\frac{\Delta}{\rho}
\partial_{\widetilde{m}}(\psi\pm i\frac{\rho}{\Delta}),
\nonumber\\
(\mp\pm)&\equiv&\widetilde{e}_{\mp}^{{\bar{m}}}A_{{\bar{m}}\pm}
\nonumber\\
&=&f_x^1(\alpha f_0^{\widetilde{m}}\pm if_1^{\widetilde{m}})
\partial_{\widetilde{m}}{\rho}.
\label{pmpmmppm}
\end{eqnarray}
Recalling (\ref{KN}), we see that they are good variables to
describe Matzner-Misner's symmetry.
%
%

We now look for canonical pairs which transform under
Matzner-Misner's SL(2,R) in a simple way.
$A_{z3}$ and $A_{z2}$ are written in terms of the variables
(\ref{pmpmmppm}) as
\begin{eqnarray}
A_{z3}&=&\frac{1}{4\widetilde{e}_3^z}\{
(++)+(-+)+(+-)+(--)
\},\nonumber\\
A_{z2}&=&\frac{1}{4i\widetilde{e}_3^z}\{
(++)+(-+)-(+-)-(--)
\}.
\label{Az32MM}
\end{eqnarray}
$A_{y2}$ can also be written as
\begin{equation}
A_{y2}=
-\frac{1}{4}\left[
\frac{\widetilde{e}_-^z}{\widetilde{e}_2^y\widetilde{e}_3^z}
\{
(++)-(+-)
\}
-
\frac{\widetilde{e}_+^z}{\widetilde{e}_2^y\widetilde{e}_3^z}
\{
(-+)-(--)
\}
\right].
\label{Ay2MM}
\end{equation}
Hence up to a total derivative we may then write
\begin{eqnarray}
&&\dot{A_{y2}}\widetilde{e}_2^y
+\dot{A_{z2}}\widetilde{e}_2^z
+\dot{A_{z3}}\widetilde{e}_3^z\nonumber\\
&=&-A_{y2}\dot{\widetilde{e}_2^y}
-A_{z2}\dot{\widetilde{e}_2^z}
-A_{z3}\dot{\widetilde{e}_3^z}\nonumber\\
&=&
-\frac{\widetilde{e}_2^y}{4\widetilde{e}_3^z}
\dot{\left(\frac{\widetilde{e}_-^z}{\widetilde{e}_2^y}
\right)}
\cdot(++)
-\frac{\widetilde{e}_2^y}{4\widetilde{e}_3^z}
\dot{\left(\frac{\widetilde{e}_+^z}{\widetilde{e}_2^y}
\right)}
\cdot(--)
-\frac{\dot{(\widetilde{e}_2^y\widetilde{e}_3^z)}}
{4\widetilde{e}_2^y\widetilde{e}_3^z}
\cdot
\{
(+-)+(-+)
\}\nonumber\\
&&+\frac{1}{2}\partial_x\widetilde{e}_1^x\cdot
\frac{\widetilde{e}_2^y}{\widetilde{e}_3^z}
\dot{\left(\frac{\widetilde{e}_2^z}{\widetilde{e}_2^y}
\right)},\label{Lform23yz}
\end{eqnarray}
where we have used
\begin{equation}
(+-)-(-+)=2i\partial_x\widetilde{e}_1^x,
\label{(+-)-(-+)}
\end{equation}
which can be shown by the relation (\ref{solvedA_y3}).
Although the first three terms of (\ref{Lform23yz})
are good canonical pairs obeying
a simple transformation rule, the last term is not because it
contains $\dot{\left(\frac{\widetilde{e}_2^z}{\widetilde{e}_2^y}
\right)}=\dot{\psi}$. If one looks at the expression of
Ashtekar's connection (\ref{AshtekarMM}), one may guess what the
good canonical variables are. It turns out that the combinations
\begin{equation}
(++)-i\widetilde{e}_1^x\frac{\widetilde{e}_2^y}{\widetilde{e}_3^z}
\partial_x\left(\frac{\widetilde{e}_+^z}{\widetilde{e}_2^y}
\right)~~
\mbox{\rm and}
{}~~
(--)+i\widetilde{e}_1^x\frac{\widetilde{e}_2^y}{\widetilde{e}_3^z}
\partial_x\left(\frac{\widetilde{e}_-^z}{\widetilde{e}_2^y}
\right),
\end{equation}
rather than $(++)$ and $(--)$ themselves, are such good variables
(if multiplied by
$-\frac{\widetilde{e}_2^y}{4\widetilde{e}_3^z})$).
By this replacement the first two terms of (\ref{Lform23yz}) read

\begin{eqnarray}
&&\mbox{(the first two terms of (\ref{Lform23yz}))}\nonumber\\
&=&
-\frac{\widetilde{e}_2^y}{4\widetilde{e}_3^z}\left[
\dot{\left(\frac{\widetilde{e}_-^z}{\widetilde{e}_2^y}
\right)}
\cdot\left\{(++)
-i\widetilde{e}_1^x\frac{\widetilde{e}_2^y}{\widetilde{e}_3^z}
\partial_x\left(\frac{\widetilde{e}_+^z}{\widetilde{e}_2^y}
\right)\right\}
+\dot{\left(\frac{\widetilde{e}_+^z}{\widetilde{e}_2^y}
\right)}
\cdot\left\{(--)
+i\widetilde{e}_1^x\frac{\widetilde{e}_2^y}{\widetilde{e}_3^z}
\partial_x\left(\frac{\widetilde{e}_-^z}{\widetilde{e}_2^y}
\right)\right\}\right]\nonumber\\
&&+\frac{1}{2}
\widetilde{e}_1^x
\left(\frac{\widetilde{e}_2^y}{\widetilde{e}_3^z}\right)^2
\left[
\dot{\left(\frac{\widetilde{e}_3^z}{\widetilde{e}_2^y}\right)}
\cdot\partial_x
\left(\frac{\widetilde{e}_2^z}{\widetilde{e}_2^y}\right)
-
\dot{\left(\frac{\widetilde{e}_2^z}{\widetilde{e}_2^y}\right)}
\cdot\partial_x
\left(\frac{\widetilde{e}_3^z}{\widetilde{e}_2^y}\right)
\right].\label{B}
\end{eqnarray}
The sum of the last term of (\ref{Lform23yz})
and the second term of (\ref{B}) simply gives
$\frac{1}{2}\dot{\widetilde{e}_1^x}\cdot
\frac{\widetilde{e}_2^y}{\widetilde{e}_3^z}
\partial_x
\left(\frac{\widetilde{e}_2^z}{\widetilde{e}_2^y}\right)$
up to a total derivative term. Thus we have succeeded to rewrite
the Liouville form as follows:
\begin{eqnarray}
&&\dot{A_{x1}}\widetilde{e}_1^x
+\dot{A_{y2}}\widetilde{e}_2^y
+\dot{A_{z2}}\widetilde{e}_2^z
+\dot{A_{z3}}\widetilde{e}_3^z\nonumber\\
&=&
\dot{
\left(
\frac{\widetilde{e}_+^z}{\widetilde{e}_2^y}
\right)
    }p_+
+
\dot{
\left(
\frac{\widetilde{e}_-^z}{\widetilde{e}_2^y}
\right)
    }p_-
+
\dot{
(
\widetilde{e}_2^y\widetilde{e}_3^z
)
    }p_{23}
+
\dot{
\widetilde{e}_1^x
    }p_1,\label{Lform}
\end{eqnarray}
where
\begin{eqnarray}
p_+&=&-\frac{\widetilde{e}_2^y}{4\widetilde{e}_3^z}
\left\{
(--)+i\widetilde{e}_1^x\frac{\widetilde{e}_2^y}{\widetilde{e}_3^z}
\partial_x\left(\frac{\widetilde{e}_-^z}{\widetilde{e}_2^y}
\right)
\right\},
\nonumber\\
p_-&=&
-\frac{\widetilde{e}_2^y}{4\widetilde{e}_3^z}
\left
\{(++)-i\widetilde{e}_1^x\frac{\widetilde{e}_2^y}{\widetilde{e}_3^z}
\partial_x\left(\frac{\widetilde{e}_+^z}{\widetilde{e}_2^y}
\right)
\right\},\nonumber\\
p_{23}&=&
-\frac{1}{4\widetilde{e}_2^y\widetilde{e}_3^z}
\{(-+)+(+-)\},\nonumber\\
p_1
&=&
-A_{x1}+\frac{\widetilde{e}_2^y}{\widetilde{e}_3^z}
\partial_x\left(\frac{\widetilde{e}_2^z}{\widetilde{e}_2^y}
\right).
\end{eqnarray}
Matzner-Misner's SL(2,R) acts on these canonical variables
in the following simple way:
\begin{eqnarray}
\left(
\frac{\widetilde{e}_+^z}{\widetilde{e}_2^y},~
p_+
\right)
&\rightarrow&
\left(
\frac{1}{i}\cdot
\frac{ia\frac{\widetilde{e}_+^z}{\widetilde{e}_2^y}+b}
{ic\frac{\widetilde{e}_+^z}{\widetilde{e}_2^y}+d},~
(ic\frac{\widetilde{e}_+^z}{\widetilde{e}_2^y}+d
)^2p_+
\right),\nonumber\\
\left(
\frac{\widetilde{e}_-^z}{\widetilde{e}_2^y},~
p_-
\right)
&\rightarrow&
\left(
-\frac{1}{i}\cdot
\frac{-ia\frac{\widetilde{e}_-^z}{\widetilde{e}_2^y}+b}
{-ic\frac{\widetilde{e}_-^z}{\widetilde{e}_2^y}+d},~
(-ic\frac{\widetilde{e}_-^z}{\widetilde{e}_2^y}+d
)^2p_-
\right),\nonumber\\
(\widetilde{e}_{2}^y\widetilde{e}_{3}^z,~
p_{23})&\rightarrow&
(\widetilde{e}_{2}^y\widetilde{e}_{3}^z,~
p_{23}),\nonumber\\
(\widetilde{e}_{1}^x,~
p_1)&\rightarrow&
(\widetilde{e}_{1}^x,~
p_1).\label{MMcanvar}
\end{eqnarray}
Obviously $\widetilde{e}_{\pm}^z/\widetilde{e}_2^y$
correspond to the Ernst potential and its complex conjugate
in the case of Ehlers' SL(2,R). The Liouville form
(\ref{Lform}) is manifestly invariant under the transformation
(\ref{MMcanvar}).

Let us rewrite again the constraints into manifestly invariant
forms by using these canonical variables. The Hamiltonian
constraint in this reduced model is given by \cite{HS}
\begin{equation}
{\cal H}~=~-2\epsilon_{1\bar{b}\bar{c}}D_xA_{{\bar{m}}\bar{b}}
\cdot
\widetilde{e}_{\bar{c}}^{{\bar{m}}}\widetilde{e}_{1}^x
{}~+2(A_{y2}A_{z3}-A_{y3}A_{z2})
\widetilde{e}_{2}^y\widetilde{e}_{3}^z, \label{HconstraintMM}
\end{equation}
where $D_x$ stands for the covariant derivative with respect to
the Ashtekar connection. The first term of (\ref{HconstraintMM})
may be written as
\begin{equation}
i\widetilde{e}_{1}^x\left[
\widetilde{e}_{-}^{{\bar{m}}}
\partial_xA_{{\bar{m}} +}
-
\widetilde{e}_{+}^{{\bar{m}}}
\partial_xA_{{\bar{m}} -}
-i A_{x1}\cdot
\left\{
(-+)+(+-)
\right\}
\right].
\end{equation}
Making use of the relations (\ref{(+-)-(-+)})
and (\ref{solvedA_y3}), one may further rewrite this as
\begin{eqnarray}
&=&\widetilde{e}_{1}^x\left[
\frac{\widetilde{e}_{2}^y}{2i\widetilde{e}_{3}^z}
\left\{
\partial_x\left(
\frac{\widetilde{e}_{-}^z}{\widetilde{e}_{2}^y}\right)
(++)
-
\partial_x\left(
\frac{\widetilde{e}_{+}^z}{\widetilde{e}_{2}^y}\right)
(--)
\right\}
-p_1\{(-+)+(+-)\}\right.\nonumber\\
&&\left.
+2\partial_x\partial_x\widetilde{e}_{1}^x
-
\partial_x\widetilde{e}_{1}^x\cdot
\frac{\partial_x(\widetilde{e}_{2}^y\widetilde{e}_{3}^z)}
{\widetilde{e}_{2}^y\widetilde{e}_{3}^z}
\right].\label{H1stterm}
\end{eqnarray}
It is easy to see that
the second term of (\ref{HconstraintMM}) is equal to
\begin{equation}
-\frac{1}{2}(++)(--)+\frac{1}{2}(-+)(+-). \label{H2ndterm}
\end{equation}
Summing up (\ref{H1stterm}) and (\ref{H2ndterm}), we end up
with the following manifestly invariant expression of ${\cal H}$:
\begin{eqnarray}
{\cal H}&=&-\frac{1}{2}\left[
\left(
\frac{4\widetilde{e}_{3}^z}{\widetilde{e}_{2}^y}
\right)^2p_+p_-
-\left(
\frac{\widetilde{e}_{1}^x\widetilde{e}_{2}^y}
{\widetilde{e}_{3}^z}
\right)^2
\partial_x
\left(
\frac{\widetilde{e}_{+}^z}
{\widetilde{e}_{2}^y}
\right)
\partial_x
\left(
\frac{\widetilde{e}_{-}^z}
{\widetilde{e}_{2}^y}
\right)
\right]
+4\widetilde{e}_{1}^x\widetilde{e}_{2}^y\widetilde{e}_{3}^z
p_{23}p_1\nonumber\\
&&+\widetilde{e}_{1}^x\left(
2\partial_x\partial_x\widetilde{e}_{1}^x
-\partial_x\widetilde{e}_{1}^x\cdot
\frac{\partial_x(\widetilde{e}_{2}^y\widetilde{e}_{3}^z)}
{\widetilde{e}_{2}^y\widetilde{e}_{3}^z}
\right)
+2(\widetilde{e}_{2}^y\widetilde{e}_{3}^z)^2 (p_{23})^2
+\frac{1}{2}(\partial_x\widetilde{e}_{1}^x)^2.
\end{eqnarray}

        For the diffeomorphism constraints,
the only remaining first class one is
\begin{eqnarray}
C_x&=&\partial\widetilde{e}_{1}^x\cdot A_{x1}-
\widetilde{e}_{\bar{a}}^{{\bar{m}}}\partial_x A_{{\bar{m}}\bar{a}}.
\end{eqnarray}
After some similar rearrangement of formulas we find
\begin{equation}
C_x~=~-\left[
\partial_x\widetilde{e}_{1}^x\cdot p_1
+\partial_x\left(
\frac{\widetilde{e}_{+}^z}{\widetilde{e}_{2}^y}
\right)\cdot p_+
+\partial_x\left(
\frac{\widetilde{e}_{-}^z}{\widetilde{e}_{2}^y}
\right)\cdot p_-
+\partial_x(
\widetilde{e}_{3}^z\widetilde{e}_{2}^y
)\cdot p_{23}
\right],
\end{equation}
which is clearly invariant under Matzner-Misner's SL(2,R)
(\ref{MMcanvar}).
\subsection{ conserved charges}
        It is also straightforward to calculate the conserved
charges for Matzner-Misner's SL(2,R).
Corresponding to the generators of sl(2,R) (\ref{SL(2,R)}),
we may in this case write down the infinitesimal Matzner-Misner's
SL(2,R) action on the canonical variables as
$h^{(MM)}, e^{(MM)}, f^{(MM)}$. For
$\widetilde{e}_{\pm}^z/\widetilde{e}_{2}^y
\equiv q_{\pm}$ we have
\begin{eqnarray}
h^{(MM)}(q_{\pm})
&=&
-2q_{\pm},\nonumber\\
e^{(MM)}(q_{\pm})
&=&\pm\frac{1}{i},\nonumber\\
f^{(MM)}(q_{\pm})
&=&\pm\frac{1}{i}
(q_{\pm})^2, \label{MMaction}
\end{eqnarray}
while
\begin{equation}
h^{(MM)}(\widetilde{e}_1^x)=
e^{(MM)}(\widetilde{e}_1^x)=
f^{(MM)}(\widetilde{e}_1^x)=
h^{(MM)}(\widetilde{e}_2^y\widetilde{e}_3^z)=
e^{(MM)}(\widetilde{e}_2^y\widetilde{e}_3^z)=
f^{(MM)}(\widetilde{e}_2^y\widetilde{e}_3^z)=0.
\end{equation}
The conserved charges read
\begin{eqnarray}
K_{h^{(MM)}}&\equiv&
\int -2\left(p_+ q_+
+ p_- q_-\right),\nonumber\\
K_{e^{(MM)}}&\equiv&\int \frac{1}{i}
\left(p_+ -p_-\right),\nonumber\\
K_{f^{(MM)}}&\equiv&\int \frac{1}{i}
\left(p_+(q_+)^2-p_-(q_-)^2\right).
\label{MMconservedcharges}
\end{eqnarray}
These expressions are completely the same as the ones in the
previous section if one replaces
$(\overline{\phi}, p_{\overline{\phi}})$ and
$(\phi,p_{\phi})$ by $(q_+,p_+)$ and $(q_-,p_-)$, respectively.
This is a consequence of Kramer-Neugebauer's transformation
which relates the two SL(2,R) symmetry in the two-Killing
reduced model \cite{KramerNeugebauer}
\footnote{Note that the conserved
charges for Ehlers' SL(2,R) are still given by
(\ref{Ehlerconservedcharges}) if the integration is performed
in the one-dimensional coset space, since the further reduction
from three to two dimensions affect only the invariant sector
for Ehlers' SL(2,R).}.

        It is not a coincidence that both of the conserved
charges (\ref{Ehlerconservedcharges}) and
(\ref{MMconservedcharges}) consist of a part of realization
of the classical $w_{\infty}$ algebra \cite{w_infty}
in terms of canonical pairs.
It is known that, in general, if one has a canonical pair
$(q,p)$ with a Poisson bracket $\{q,~p\}=1$, one can
realize the classical $w_{\infty}$ algebra by assigning
\begin{equation}
{\rm W}_{n}^{(l)}=p^{l-1}q^{n+l-1}~~~(l\geq 1,~n\geq -l+1).
\end{equation}
Their Poisson bracket satisfies the commutation relation
of the $w_{\infty}$ algebra
\begin{equation}
\{{\rm W}_{n}^{(l)},~{\rm W}_{m}^{(k)}\}
=((k-1)n-(l-1)m){\rm W}_{n+m}^{(l+k-2)},
\end{equation}
known as an algebra of the area-preserving diffeomorphism.
Indeed, the canonical transformation is by definition a
transformation that preserves the area of $(q,p)$ phase space.

${\rm W}_{n}^{(l)}$ is then the generating function of the
canonical transformation. This algebra contains ``half'' of the
Virasoro (Witt) subalgebra $(L_n\geq -1)$
generated by $L_n={\rm W}_{n}^{(2)}$, and this
Virasoro algebra further contains therein the sl(2,R)
generated by $\{L_{-1}=p, L_0=pq, L_{1}=pq^2\}$. Both
conserved charges (\ref{Ehlerconservedcharges}) and
(\ref{MMconservedcharges}) are
diagonal sums of sl(2,R) generators
made out of two sets of canonical pairs.
Both for Ehlers' and Matzner-Misner's case, we have succeeded
to take two canonical pairs, only on which the sl(2,R) in
question act as canonical transformations. Therefore, the
algebra generated by the symmetry charges is necessarily a
subalgebra of the canonical transformation on these two
canonical pairs, and this subalgebra contains the diagonal
$w_{\infty}$ algebra as a special case. What we have observed
here is that the diagonal sl(2,R) in this diagonal
$w_{\infty}$ realizes the symmetry algebra.

\subsection{GL(2,R) in
[19]
and the central extension in the Geroch group}
In ref. \cite{HS} a set of conserved quantities was found
in the two-Killing reduced Einstein gravity. They are given by
in our notation
\begin{equation}
K_{\bar{n}}^{\bar{m}}
\equiv\int\widetilde{e}_{\bar{a}}^{\bar{m}}A_{\bar{n}\bar{a}}
\label{HSconservedcharges}
\end{equation}
($\bar{m},\bar{n}=y,z$, and $\bar{a}$ runs over $\{2,3\}$).
The Poisson brackets between two of
these quantities form the $\mbox{\rm gl(2,R)}\sim\mbox{\rm sl(2,R)}
\oplus{\rm R}$. Let us examine to which symmetry this gl(2,R)
corresponds in the Geroch group.
In fact, these conserved charges in the sl(2,R)
sector are precisely the ones associated with Matzner-Misner's
symmetry \footnote{This fact has been already pointed out by
H.Nicolai \cite{Nicolai_integrable}.}.
Indeed, the three generators
\begin{equation}
\{K_{z}^{z}-K_{y}^{y}, K_{z}^{y},K_{y}^{z}\}
\label{HSsl2R}
\end{equation}
constitute the
sl(2,R), corresponding to $\{h,e,f\}$, respectively, while the
trace $K_{z}^{z}+K_{y}^{y}$ commutes with any of these elements.
Using
(\ref{solvedA_y3}), (\ref{Ay2MM}), (\ref{Az32MM}) and
(\ref{(+-)-(-+)}) and integrating by parts, it is
straightforward to check that the conserved quantities
(\ref{HSsl2R}) precisely reproduce the conserved charges
(\ref{MMconservedcharges}).

The rather complicated look of the conserved charges
(\ref{MMconservedcharges}) is in fact an artifact of the
gauge-fixing (\ref{MMparametrization}). Indeed,
Matzner-Misner's symmetry can be seen as a symmetry that mixes
the $y$ and $z$ indices. If we do not fix
$\widetilde{e}_3^y=0$ but restore the full
$\widetilde{e}_{\bar{a}}^{\bar{m}}$ ($\bar{m}=y,z$; $\bar{a}=2,3$),
such a variation can be written as
\begin{equation}
\delta(\widetilde{e}_{\bar{a}}^{\bar{m}})
=\widetilde{e}_{\bar{a}}^{\bar{n}}X_{\bar{n}}^{\bar{m}}
\label{Xvariation}
\end{equation}
for some $X_{\bar{n}}^{\bar{m}}\in \mbox{\rm sl(2,R)}$. If we
take $X_{\bar{n}}^{\bar{m}}=-h,e,f$ defined in (\ref{SL(2,R)})
(The minus sign for $h$ is because the variation is a right
action on $\widetilde{e}_{\bar{a}}^{\bar{n}}$.),
the conserved charges associated with the invariance under these
variations, which can be verified in the expressions of
the constraints in ref.\cite{HS}
\footnote{
{}From the fact that all the indices $\bar{n}$ of
$\widetilde{e}_{\bar{a}}^{\bar{n}}$ (corresponding to the index
``$\alpha$'' in ref.\cite{HS}'s notation) are contracted by those
of $A_{\bar{n}\bar{b}}$.},
are nothing but the conserved
quantities (\ref{HSconservedcharges}) obtained there.

What role does the trace of the gl(2,R) play in the Geroch group,
then? To answer this question, let us first recall how the two
sl(2,R) Lie algebras are combined to give the affine Kac-Moody
algebra \cite{Kac} $\rm\widehat{sl}(2,R)$.
The affine Kac-Moody algebra $\rm\widehat{sl}(2,R)$ is defined by
the following commutation relations:
\begin{eqnarray}
[H_n, E_m]&=&2E_{n+m},\nonumber\\
{[}H_n, F_m{]}&=&-2F_{n+m},\nonumber\\
{[}E_n, F_m{]}&=&nk\delta_{n+m,0}+H_{n+m},\nonumber\\
{[}H_n, H_m{]}={[}E_n, E_m{]}&=&{[}F_n, F_m{]}=0,
\end{eqnarray}
where $k\in{\rm R}$ belongs to the center of this algebra,
and $n,m\in{\rm Z}$. This contains following two sl(2,R)
subalgebras:
\begin{eqnarray}
&&
{[}H_0, E_0{]}=2E_{0},~
{[}H_0, F_0{]}=-2F_{0},~
{[}E_0, F_0{]}=H_{0},\nonumber\\
&&
{[}k-H_0, F_1{]}=2F_{1},~
{[}k-H_0, E_{-1}{]}=-2E_{-1},~
{[}F_1, E_{-1}{]}=k-H_{0}.
\label{twosl2R}
\end{eqnarray}
Conversely, let us assume that the two sets of sl(2,R)
generators $\{h_i, e_i, f_i\}$ $(i=0,1)$ satisfy
\begin{equation}
{[}h_i, h_j{]}=0,~
{[}h_i, e_j{]}=A_{ij}e_j,~
{[}h_i, f_j{]}=-A_{ij}f_j,~
{[}e_i, f_j{]}=\delta_{ij}h_j
\label{Serre1}
\end{equation}
and
\begin{equation}
({\rm ad}e_i)^{1-A_{ij}}(e_j)=0,~
({\rm ad}f_i)^{1-A_{ij}}(f_j)=0
\label{Serre2}
\end{equation}
for $i\neq j$, where the Cartan matrix $A_{ij}$ reads
in this case
\begin{equation}
A_{ij}=\left(
\begin{array}{rr}2&-2\\
-2&2\end{array}
\right).
\end{equation}
One may define the $\rm\widehat{sl}(2,R)$ algebra as
a Lie algebra generated by any successive multiplication
of commutators satisfying (\ref{Serre1})(\ref{Serre2}).
The set of (\ref{Serre1})(\ref{Serre2}) is called
the Serre relation.

It has been shown that (the Lie algebra of) the Geroch group
can be obtained by making use of two sl(2,R) algebras, one of
which is Ehlers' and the other of which is Matzner-Misner's,
as the ones required in the Serre relation.
One of the interesting features of the Geroch group is that
it realizes the central-extended $\rm\widehat{sl}(2,R)$
algebra even at the classical level already. Usually,
a central term such as in (\ref{twosl2R}) arises as
a consequence of anomaly in quantum field theory.
Here the situation is different;
identifying the two sl(2,R)'s as the ones corresponding to the
two simple roots of the $\rm\widehat{sl}(2,R)$,
the central element\footnote{The Virasoro algebra generated
by the Sugawara form of this $\rm\widehat{sl}(2,R)$ should not
be confused with the Virasoro subalgebra of the $w_{\infty}$
in the last subsection. Obviously they are different things.
In particular,  the latter has only the half of generators
$L_n$ for $n\geq -1$ and hence the ``central extension'' has no
meaning.} $k=H_0+(k-H_0)=h_0+h_1$ then acts
non-trivially on the fields, although we are still considering
classical gravity theory.

Let us now see how this central element acts on the Ashtekar
variables. We may identify $h_0=h^{(E)}$ and $h_1=h^{(MM)}$.
For $h^{(E)}$, we have seen in sect.3 that it acts on
$\overline{\phi}, \phi, \widetilde{e}_{\alpha}^{\mu}$ as,
respectively,
\begin{equation}
h^{(E)}(\overline{\phi})=-2\overline{\phi},~
h^{(E)}(\phi)=-2\phi,~
h^{(E)}(\widetilde{e}_{\alpha}^{\mu})=0.
\end{equation}
This means that
\begin{equation}
h^{(E)}(\Delta)=-2\Delta,~h^{(E)}(B)=-2B.
\end{equation}
Hence, in the present two-Killing reduced model,
it acts on the parameters in (\ref{MMparametrization}) as
\begin{equation}
h^{(E)}(\Delta)=-2\Delta,~
h^{(E)}(\psi)=+2\psi,~
h^{(E)}(f_x^1)=
h^{(E)}(\rho)=0,
\end{equation}
where we have used the relation (\ref{duality2}).
We also know, on the other hand, the $h^{(MM)}$ action
(\ref{MMaction}) on the parameters
(\ref{MMparametrization}), which reads
\begin{equation}
h^{(MM)}(\Delta)=+2\Delta,~
h^{(MM)}(\psi)=-2\psi,~
h^{(MM)}(f_x^1)=+f_x^1,~
h^{(MM)}(\rho)=0.
\end{equation}
The action of the central element $k$ is thus given by
\begin{equation}
k(f_x^1)=+f_x^1,~
k(\Delta)=
k(\psi)=
k(\rho)=0.
\end{equation}
Therefore it causes a scale transformation only on $f_x^1$
without doing anything on the other parameters. This confirms
the known fact that the central element acts as a rescaling
on the conformal factor of the zweibein for the ``uncompactified''
sector. On the Ashtekar variables we have
\begin{equation}
k(\widetilde{e}_{\bar{a}}^{\bar{m}})
=\widetilde{e}_{\bar{a}}^{\bar{m}},~
k(A_{\bar{m}\bar{a}})
=-A_{\bar{m}\bar{a}},~
k(\widetilde{e}_{1}^{x})
=k(A_{x1})=0,
\end{equation}
where $(\bar{m},\bar{a})\neq(y,3)$ in our gauge.
We thus find that the central element $k$ acts as a scale
transformation on the ``compactified'' sector
($\widetilde{e}_{\bar{a}}^{\bar{m}},
A_{\bar{m}\bar{a}})$,
while keeping the ``uncompactified'' sector
$(\widetilde{e}_{1}^{x},
A_{x1})$ invariant.

Let us go back to the question of the role of the center in the
GL(2,R). From the analysis in the previous paragraphs,
it is now obvious that the action of the trace of gl(2,R)
(\ref{HSconservedcharges}) is precisely the same as that of
the central element of the Geroch group. Indeed, the variation
on $\widetilde{e}_{\bar{a}}^{\bar{m}}$'s from the action of the
trace of gl(2,R) can be achieved by taking $X_{\bar{n}}^{\bar{m}}$
in (\ref{Xvariation}) to be the identity matrix.

We can show that $K_y^y+K_z^z$ acts as a rescaling on
$\widetilde{e}_{\bar{a}}^{\bar{m}}$ directly also as follows.
Writing
\begin{eqnarray}
\widetilde{e}_{\bar{a}}^{y}A_{y\bar{a}}
&=&\frac{1}{2(a+b)}\left[
-b(++)+a(-+)+b(+-)-a(--)
\right],\nonumber\\
\widetilde{e}_{\bar{a}}^{z}A_{z\bar{a}}
&=&\frac{1}{2(a+b)}\left[
b(++)+a(-+)+b(+-)+a(--)
\right],
\end{eqnarray}
where $a=\widetilde{e}_+^{z}$ and $b=\widetilde{e}_-^{z}$,
we find
\begin{eqnarray}
K_y^y+K_z^z&=&\frac{1}{2}\int[(-+)+(+-)]\nonumber\\
&=&-2\int p_{23}q_{23},
\end{eqnarray}
where we set
$q_{23}\equiv\widetilde{e}_2^{y}\widetilde{e}_3^{z}$.
On the other hand, all $\widetilde{e}_2^{y}$,
$\widetilde{e}_2^{z}$ and $\widetilde{e}_3^{z}$ are
functions of $q_{\pm}$ and $q_{23}$, and in particular
they are proportional to $\sqrt{q_{23}}$. Hence, the fact that
$K_y^y+K_z^z$ scales $\widetilde{e}_2^{y}$,
$\widetilde{e}_2^{z}$ and $\widetilde{e}_3^{z}$ while leaves
$\widetilde{e}_1^{x}$ unchanged immediately follows from
the following equations:
\begin{eqnarray}
\{\sqrt{q_{23}},~-2\int p_{23}q_{23}\}&=&\sqrt{q_{23}},\nonumber\\
\{\widetilde{e}_1^{x},~-2\int p_{23}q_{23}\}&=&0.
\end{eqnarray}

To summarize
what we have shown in this subsection, the (finite, non-affine)
GL(2,R) symmetry found in ref.\cite{HS} is indeed {\em a part}
of the Geroch group, where the SL(2,R) and the trace sector
correspond to the Matzner-Misner's SL(2,R) and the central
element of the Geroch group, respectively.

\subsection{GL(2,R) loop algebra?}
Finally, let us discuss the relation between the Geroch group
and the GL(2,R) loop algebra constructed in ref.\cite{HS},
where the following
operators are considered in the loop representation:
\begin{equation}
L[\phi]\equiv\int\phi_{\bar{m}}^{\bar{n}}
\widetilde{f}^{\bar{m}}
\frac{\delta}{\delta\widetilde{f}^{\bar{n}}}.
\label{HSlooprepresentation}
\end{equation}
The $x$-coordinate is compactified into $\rm S^1$,
$0\leq x\leq 2\pi$.
$\widetilde{f}^{\bar{m}}$ are some two densities
and $\phi_{\bar{n}}^{\bar{m}}$ are any functions on
this $\rm S^1$.
$L[\delta_{\bar{n}}\delta^{\bar{m}}]$ reduces to the generators
of GL(2,R) in the last subsection, where
$\delta_{\bar{n}}\delta^{\bar{m}}$ denotes a $2\times 2$ matrix
in which only the $(\bar{n},\bar{m})$ component is 1 and
otherwise 0.

It was speculated \cite{HS} that the
loop algebra generated by (\ref{HSlooprepresentation}) might
be related to the Geroch group.
As we have shown, the GL(2,R)
algebra generated by $K_{\bar{n}}^{\bar{m}}$ consists of
Matzner-Misner's SL(2,R) and the central element of Geroch
group. Hence the GL(2,R) loop algebra indeed includes
Matzner-Misner's SL(2,R). Does this
GL(2,R) loop algebra also includes Ehlers' SL(2,R)?

In fact, this is not the case. This can be seen as follows.
First, let us examine the Serre generators
among (\ref{HSlooprepresentation}). The basis of the
$\rm\widehat{gl}(2,R)$ are obtained by expanding the function
$\phi_{\bar{n}}^{\bar{m}}$ in (\ref{HSlooprepresentation})
in terms of the Fourier modes
\begin{equation}
L[e^{ikx}\delta_{\bar{n}}\delta^{\bar{m}}], ~k\in{\rm Z}.
\label{HSbasis}
\end{equation}
The Serre generators of its $\rm\widehat{sl}(2,R)$ sector are
\begin{equation}
\{L[\delta_{z}\delta^{z}]
-L[\delta_{y}\delta^{y}],
L[\delta_{y}\delta^{z}],
L[\delta_{z}\delta^{y}]\}~~\mbox{and}~~
\{-L[\delta_{z}\delta^{z}]
+L[\delta_{y}\delta^{y}],
L[e^{ix}\delta_{z}\delta^{y}],
L[e^{-ix}\delta_{y}\delta^{z}]
\}.
\end{equation}
The first set is Matzner-Misner's SL(2,R) as we saw in the last
subsection, while the second one does {\em not} correspond to
the conserved charges for Ehlers' SL(2,R)
(\ref{Ehlerconservedcharges}) which acts {\em non-locally}
on the Ashtekar variables.

Another reason to believe the
absence of Ehlers' SL(2,R) is that the $\rm\widehat{sl}(2,R)$
defined by the traceless generators of
(\ref{HSbasis}) has vanishing central charge.
Basically, the affine Kac-Moody algebra constructed as a loop
algebra has no central term until
a cocycle term is introduced \cite{loopgroup}.
Here, as we have seen in the last subsection, the central
element of the Geroch group is already included
in the trace $L[\delta_{z}\delta^{z}]+L[\delta_{y}\delta^{y}]$,
but such a term never results from any commutators between the
two generators of (\ref{HSbasis}).
We are thus led to the conclusion
that the GL(2,R) loop algebra generated by the operators
(\ref{HSlooprepresentation}) is not the same as the Lie algebra
of the Geroch group itself, but contains only Matzner-Misner's
SL(2,R) and the central element of the Geroch group as its
GL(2,R) subalgebra.

\section{Conclusion and comment}

We have studied the realization of the Geroch group in the
Ashtekar formulation in this paper. Our first
observation was the relation between the Ashtekar connection
and the Ernst potential. In the history of the gravity theory,
the discovery of the former has brought us a chance to construct
a consistent quantum theory, where its complexification is
unavoidable in order to achieve the simplification of the
constrains. On the other hand, the latter,
introduced long before the discovery of the former,
plays a central pole in the integrable Ernst equation,
where its complex nature is the one inherited from the
complex structure of the target space of the sigma model.
There seem no reasons that they are
necessarily related, and that makes this coincidence
interesting.

We have constructed for each case of Ehlers' and Matzner-Misner's
SL(2,R) a set of canonical variables that realize canonically
either of them, but could not find the one that realizes both
at the same time. To realize the first one, we are forced to
use canonical variables which are non-local with respect to
the original ones. Difficulties to realize canonically the
full Geroch group may be guessed from a known example on the
canonical realization of the non-local symmetry in a
chiral model \cite{chiral}, which is much simpler than
the present one.

We have shown that the action of the GL(2,R) charges constructed
in ref. \cite{HS} corresponds to a subgroup of the Geroch group,
i.e. the product of Matzner-Misner's SL(2,R) and the center of
the Geroch group.  We have further examined whether or not
their GL(2,R) loop algebra contains Ehlers' SL(2,R), but it does
not. This can be seen either by comparing the Serre generators
with the Ehlers' SL(2,R) symmetry charges, or by noticing
the absence of the central term in their loop algebra.
Therefore their loop algebra is not the Lie algebra
of the Geroch group itself, but something else.

Finally we would like to comment on the recent argument that
the two-Killing reduced model has the same linear system as
the one for a flat-space chiral model \cite{H}. However, the
derivation is based on an unusual gauge-choice, which
in fact can not be achieved generically. The author of \cite{H}
starts from a block-diagonal dreibein, which is written in our
notation
\begin{equation}
e_m^a=\left(
\begin{array}{cc}
e_x^1&\\
&\bar{e}_{\bar{m}}^{\bar{a}}
\end{array}
\right)~,
\end{equation}
or equivalently
\begin{equation}
\widetilde{e}_a^m=\left(
\begin{array}{cc}
\bar{e}&\\
&e_x^1\bar{e}\bar{e}_{\bar{a}}^{\bar{m}}
\end{array}
\right)~
\end{equation}
$(\bar{m}=y,z;~ \bar{a}=2,3;~
\bar{e}=\mbox{\rm det}\bar{e}_{\bar{m}}^{\bar{a}})$.
$A_{x\bar{a}}$ and $A_{\bar{m}1}$ are also set to zero
similarly to Sec.IV. At this stage we still have three
gauge degrees of freedom generated by the first class
constraints ${\cal H}$, ${\cal C}_x$ and ${\cal G}^1$.
By adopting the gauge-fixing conditions
\begin{eqnarray}
A_{x1}&=&0,\nonumber\\
J~&\equiv&\epsilon^{\bar{a}\bar{b}}
A_{\bar{m}\bar{a}}
\widetilde{e}_{\bar{b}}^{\bar{m}}
=0,\nonumber\\
K~&\equiv&
A_{\bar{m}\bar{a}}
\widetilde{e}_{\bar{a}}^{\bar{m}}
=\mbox{\rm const.},
\end{eqnarray}
which means in consequence
\begin{eqnarray}
\widetilde{e}_1^x&=&\bar{e}=\mbox{\rm const.},\nonumber\\
N~&=&\mbox{\rm const.},\nonumber\\
N^x&=&\mbox{\rm const.}, \label{gauge_choice}
\end{eqnarray}
the author of \cite{H} was led to a linear system of the
flat-space SL(2,R) chiral model. However, in reality, the
gauge-choice (\ref{gauge_choice}) can not be achieved generically.
Indeed, the Lorentz rotation generated by  ${\cal G}^1$
can not affect the determinant $\bar{e}$, the lapse $N$
nor the shift $N^x$. Hence we may employ only the other
two gauge degrees of freedom to fix three independent elements
of the vierbein, which is not be achievable in general
unless the spacetime is flat from the beginning.
Therefore the argument of ref.\cite{H} does not show that
the two-Killing reduced model is equivalent to a flat space
chiral model.

\acknowledgments
I am deeply grateful to H. Nicolai for arising my interest in
this subject, and also for stimulating discussions,
useful comments and reading the manuscript.
I would like to thank D. Korotkin, H. -J. Matschull and J. A.
Teschner for discussions. This work was supported by the
Alexander von Humboldt Foundation.

\end{document}